\documentclass[10pt,pra,aps,twocolumn,longbibliography,floatfix]{revtex4-2}
\usepackage{times}
\usepackage{graphicx}
\usepackage{epsfig}
\usepackage{amsfonts}
\usepackage{comment}
\usepackage{amsmath}
\usepackage{amssymb}
\usepackage{bbold}
\usepackage{svg}
\usepackage{color}
\usepackage{dsfont}
\usepackage{multirow}
\usepackage[utf8]{inputenc}
\usepackage[T1]{fontenc}
\usepackage{appendix}
\usepackage{mathrsfs}
\usepackage{latexsym}
\usepackage{physics}
\usepackage{epigraph}
\def\ff{\text{f}}

\usepackage[colorlinks=true,linkcolor=blue,citecolor=blue,urlcolor=blue]{hyperref}

\begin{document}

\title{Vibrational parametric arrays with trapped ions: non-Hermitian topological phases and quantum sensing}

\author{Miguel Clavero-Rubio}
\email{miguel.clavero@iff.csic.es}
\affiliation{Institute of Fundamental Physics IFF-CSIC, Calle Serrano 113b, 28006 Madrid, Spain.}
\author{Tom\'as Ramos}
\affiliation{Institute of Fundamental Physics IFF-CSIC, Calle Serrano 113b, 28006 Madrid, Spain.}
\author{Diego Porras}
\email{diego.porras@csic.es}
\affiliation{Institute of Fundamental Physics IFF-CSIC, Calle Serrano 113b, 28006 Madrid, Spain.}

\date{\today}

\begin{abstract}
We consider a linear array of trapped ions subjected to local parametric modulation of the trapping potential and continuous laser cooling. 
In our model, the phase of the parametric modulation varies linearly along the array, breaking time-reversal symmetry and inducing non-trivial topological effects. 
The linear response to an external force is investigated with the Green's function formalism. 
We predict the appearance of topological amplification regimes in which the trapped ion array behaves as a directional amplifier of vibrational excitations. 
The emergence of topological phases is determined by a winding number related to non-Hermitian point-gap topology. 
Beyond its fundamental interests as a topological driven-dissipative system, our setup can be used for quantum sensing of ultra-weak forces and electric fields. 
We consider a scheme in which a trapped ion at one edge of the array acts as a sensor of an ultra-weak force, and the vibrational signal gets amplified towards the last trapped ion, which acts as a detector. 
We consider arrays of 2-30 $^{25}$Mg$^+$ ions, assuming that the detector ion's displacement is measured via fluorescence with a spatial resolution of 200-500 nm, and predict sensitivities as small as 1 yN $\cdot$ Hz$^{-1/2}$. Our system has the advantage that the detected force frequency can be tuned by adjusting the frequency of the periodic drive.
\end{abstract}

\maketitle

\section{Introduction}
\label{introduction}
Trapped ion setups are one of the most advanced experimental platforms for the manipulation of quantum matter. 
In trapped ion crystals, internal states and vibrational modes can be controlled and measured with great precision by means of lasers and magnetic fields \cite{Leibfried03rmp}.
This remarkable degree of control has led to many successful applications in the quantum simulation of quantum many-body models \cite{blatt2012quantum,schneider2012experimental,monroe21review}.
%
In addition to coherent interactions, dissipation can be controlled, for example, using laser cooling \cite{cirac1992laser}.
The trapped ion experimental toolbox has also found applications in the quantum sensing of electric fields and ultra-weak forces \cite{biercuk2010ultrasensitive, ivanov2020steady, bollinger_dm}. Trapped ions are, thus, ideal setups for exploring the interplay between many-body physics and metrological applications. 
%

One of the most promising research avenues in this field is the implementation of bosonic quantum models using vibrational modes. Phonons in trapped ion crystals can hop between ions in much the same way as bosons in an optical lattice \cite{porras2004bose,Deng2008pra,Haze12pra,Toyoda2015nat}. 
Floquet engineering can imprint complex phonon hopping terms and synthetic gauge fields, as proposed in \cite{bermudez_synthetic_2011} and experimentally demonstrated in \cite{kiefer2019floquet}, leading to the quantum simulation of phononic topological phases. Non-linearities and phonon interactions can be induced, for example, by coupling vibrational modes to the internal, spin degrees of freedom \cite{Porras2008pra, Ivanov2009pra, Monroe2018prl,OhiraQST2021,Katz2023prl}. 
Vibrational parametric terms, which are the bosonic counterparts of pairing terms in superconductors, can also be implemented in trapped ion setups \cite{parametric2017prl,Burd2021,Leibfried2024prx,Bazavan2024commphys}.
Finally, laser cooling can be used to induce dissipation in the form of phonon loss (laser cooling) or phonon gain (heating)  \cite{cirac1992laser}, thus opening exciting possibilities for the exploration of non-equilibrium phases \cite{Bermudez2013prl}.  

The interplay between dissipation and topology in bosonic systems gives rise to exciting effects that may be accessed with trapped ions.
For example, non-Hermitian topology in driven-dissipative bosonic chains can induce topological amplification phases in which the system becomes a directional amplifier of incoming radiation \cite{porras2019topological, wanjura2020topological,ramos2021topological,gomez2023driven}. 
The appearance of topological phases typically requires the interplay of dissipation, parametric couplings, and breaking of the time-reversal symmetry. 
This is the case, for instance, of the bosonic Kitaev chain \cite{mcdonald2018phase}, recently implemented with superconducting circuits \cite{busnaina2024quantum} and optomechanical setups \cite{slim2024optomechanical}. 
Reaching the quantum limit in large bosonic driven-dissipative systems can be challenging, especially in vibrational or opto-mechanical systems.
Here, trapped ions offer an advantage since laser cooling techniques enable us to work close to the vibrational ground state. Topological dissipative systems are fascinating in this limit, given possible applications, such as amplification or sensing \cite{gomez2023driven}. 

In this work, we propose and theoretically analyze a trapped ion setup subjected to a parametric modulation of the local trapping frequency that induces a topological amplification regime. 
In our proposal, the phase of the parametric modulation phase varies linearly with the position of the ions. 
This is essential for breaking time-reversal symmetry and inducing driven-dissipative topological regimes \cite{gomez2023driven}.
We present a theoretical description of the linear response to an external force in terms of a topological Green's function formalism \cite{ramos2021topological,gomez2023driven}. 
Topological amplification results in the exponential enhancement of a highly non-reciprocal linear response. 
Furthermore, we show that the system's response strongly depends on the frequency of the external force, and we identify optimal frequency values for amplification. We also investigate the steady-state properties of the chain, showing that even in the absence of an external perturbation, topological effects can be detected in properties such as phonon number and phonon correlations.

We propose the application of the trapped-ion parametric chain in the detection of ultra-weak forces. In our scheme, the first ion acts as a sensor upon which a weak force acts. The vibrational signal is amplified along the chain, and it is detected by measuring the position of the last ion, which acts as a detector. We analyze the signal-to-noise ratio and the typical measurement times. We show that topological amplification leads to sensitivities as low as 1 yN $\cdot$ Hz$^{-1/2}$. Our scheme has the advantage that the signal is detected by measuring the amplified displacement of the last ion, which can be as large as $0.5$ $\mu$m, and it is thus directly detectable by fluorescence measurements. Furthermore, the detection frequency is tunable by adjusting the frequency of the periodic drive.

The article is structured as follows. 
In Section \ref{section2}, we present the model interactions, including coherent phonon hopping, parametric driving of the trapping potential, local dissipation, and an external coherent field. Section \ref{section3} is dedicated to the theoretical description of the parametric trapped ion chain.  We present the linear response in terms of Green's functions and the steady-state properties, together with the criterion for stability.  In Section \ref{section5}, we study our system under the lens of topological amplification theory and calculate a non-equilibrium phase diagram with topological phase transitions between topological and trivial steady-states. In Section \ref{section6}, we establish quantum sensing as an advantageous application of these topological phases and characterize the efficiency of the systems in detecting ultra-weak forces. We finish with conclusions in Section \ref{sec.conclusions}.

\section{Dissipative parametric interactions with trapped ions}  
\label{section2}
In this section, we deduce the vibrational Hamiltonian for a trapped ion chain in the presence of parametric driving of the trapping frequencies and phonon decay processes induced by laser cooling.

\subsection{Vibrational Hamiltonian for ions in microtraps}

\begin{figure}[h]
    \centering
    \includegraphics[width=0.5\textwidth]{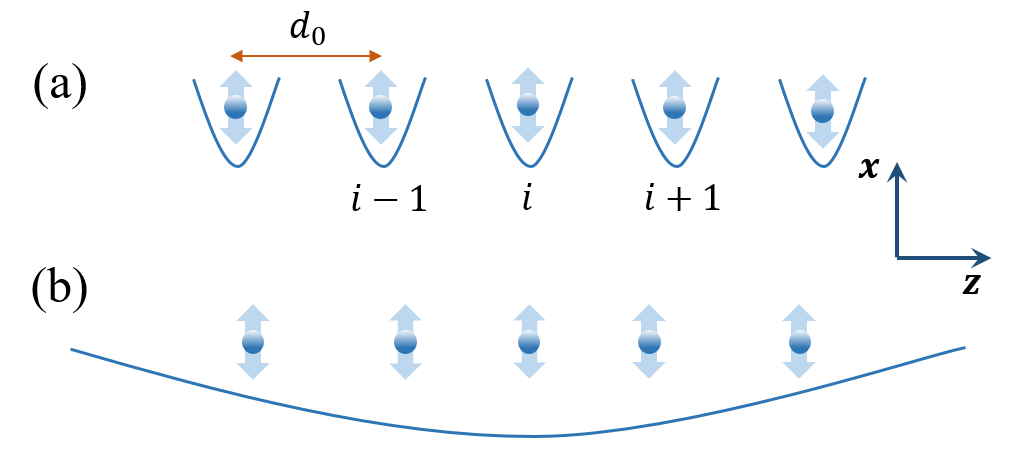}
    \caption{Schematic of the trapped-ion chain. We consider the motion along the $x$-axis, perpendicular to the linear chain oriented along the $z$-axis. (a) Individual microtraps. (b) Ion chain in a linear trap.}
    \label{schemearray2}
\end{figure}

We consider a linear array of $N$ ions with mass $m$ and charge $e$, placed in a linear electromagnetic trap or in an array of individual microtraps. 
We will assume that the motion of the ions is excited and measured only in the $x$-direction, perpendicular to the linear array, which is placed in the $z$-direction (see Fig. \ref{schemearray2}). 

The position of the ions is
\begin{equation}
{\bf r}_i = x_i {\bf x} + d_i {\bf z},
\end{equation}
where $x_i$ are displacement operators of the $N$ ions around the equilibrium position, and $d_i$  are the equilibrium positions along the trapping axis. 
We assume that ions are equally spaced by a constant distance, $d_0$. This can be achieved by pinning the ions' positions with local electrodes or using anharmonic axial trapping potentials \cite{Pagano2019}. 
Other vibrational configurations are also possible, for example, involving both $x$ and $y$ vibrations.
The trapping Hamiltonian $H_0$ is 
\begin{equation}
    H_{\rm 0} = \sum_{i} \left( \frac{p_i^2}{2 m} + \frac{1}{2} m \omega_{\rm t}^2 x_i^2 \right),
    \label{H.H0}
\end{equation}
where $p_i$  account for the momentum of each ion and $\omega_{\rm t}$ is the radial trapping frequency, which we assume to be constant. Indexes $i$, $j$, representing ion site always run from $1$ to $N$ along the paper.

The ions are coupled by the Coulomb interaction through the Hamiltonian term $H_{\rm C}$, given by (in Gaussian units)
\begin{equation}
    H_{\rm C} = \frac{e^2}{2}\sum_{\substack{i,j \\ (i \neq j)}} \frac{1}{|{\bf r}_i - {\bf r}_j|}.
\end{equation}
We expand the Coulomb interaction up to second order \cite{james1997quantum} in the ion displacements $x_i$,
\begin{eqnarray}
H_{\rm C} 
&\simeq& - \frac{e^2}{4} \sum_{\substack{i,j \\ (i \neq j)}} \frac{1}{|d_i - d_j|^3} \left( x_i - x_j \right)^2 
\nonumber \\
&=& \sum_i \bar{V}_{i} x_i^2 
+ \frac{1}{2} \sum_{\substack{i,j \\ (i \neq j)}} V_{ij} x_i x_j,
\label{H.quad}
\end{eqnarray}
where 
\begin{eqnarray}
\bar{V}_{i}  &=& - \frac{e^2}{2}  \sum_{\substack{j \\ (j \neq i)}} \frac{1}{|d_i - d_j|^3} , \nonumber \\
V_{ij} &=&   \frac{e^2}{|d_i - d_j|^3} .
\end{eqnarray}

We proceed now to the quantization of the ions' $x,p$-coordinates. Here and throughout the manuscript, we set $\hbar=1$, leading to
\begin{eqnarray}
          x_i & = &\sqrt{\frac{1}{2 m \omega_{\rm t} }}\left(a^\dagger_i + a_i \right), \\
          p_i & = &i \sqrt{\frac{m \omega_{\rm t}}{2}} \left(a^\dagger_i - a_i \right) .
\end{eqnarray}
The total vibrational Hamiltonian 
$H_{\rm v} = H_0 + H_{\rm C}$ can be rewritten now in terms of bosonic creation-annihilation operators $a_i^\dagger$, $a_j$, as
\begin{eqnarray}
H_{\rm v} &\approx& \sum_i \omega_{{\rm t},i} a^\dagger_i a_i 
+ \frac{1}{2} \sum_{\substack{i,j \\ (i \neq j)}}\frac{J_{\rm c}}{|i-j|^3}  \left( a^\dagger_i a_j  + a_i a^\dagger_j \right) ,
\end{eqnarray}
where the Coulomb coupling strength is given by
\begin{equation}
J_{\rm c} = \frac{e^2}{2 m \omega_{\rm t} d_0^3} .
\end{equation}

We have neglected cross-terms $a_i a_j$ and $a_i^\dagger a_j^\dagger$, in a rotating wave approximation, which is well justified in the usual limit $\omega_{\rm t} \gg J_{\rm c}$. 
Finally, the frequencies $\omega_{{\rm t},i}$ become site dependent, since they get a correction from the mean Coulomb interactions,
\begin{equation}
\omega_{{\rm t},i}^2 = \omega_{\rm t}^2 + \frac{1}{2 m \omega_{\rm t}} \bar{V}_{i} .
\end{equation}
In the following we neglect these site-dependent corrections, which are typically small \cite{porras2004bose},  and approximate $\omega_{{\rm t},i} \approx \omega_{\rm t}$.
\subsection{Parametric driving of the trapping potential}
We consider parametric driving terms induced by a time-periodic potential 
\begin{equation}
H_{\rm d}(t) = K \sum_i 
\cos(2 \omega_{\rm d} t - 2\phi_i) x_i^2,
\label{eq:parametric.drive}
\end{equation}
where $\omega_{\rm d}$ is the driving frequency and $\phi_i$ is a phase that depends on the site, something that is crucial to break the time-reversal symmetry and obtain nontrivial topological regimes. 
A local parametric drive can be implemented by applying localized electrodes \cite{parametric2017prl, Leibfried2024prx} or by using optical forces 
\cite{Katz2023prl}.
We assume that the driving frequency is close to the trapping frequency, up to a detuning $\Delta$, 
\begin{equation}
\omega_{\rm d} = \omega_{\rm t} - \Delta.
\end{equation}
with $\Delta$ small, $\Delta \ll \omega_{\rm d}, \omega_{\rm t}$. We define the parametric driving amplitude 
$g = K x_0^2 = K / 2 m \omega_{\rm t}$, which we also assume small, $g \ll \omega_{\rm t}$. 

The analysis of our system is simpler in a rotating frame with respect to the frequency of the parametric drive, 
\begin{equation}
a_i \to a_i e^{-i \omega_{\rm d} t} .
\label{rot.frame}
\end{equation}
In this frame, and after a rotating wave approximation, we can approximate the total Hamiltonian $H = H_0 + H_{\rm C} + H_{\rm d}(t)$ by 
\begin{equation}
H \approx \Delta \sum_i a^\dagger_i a_i  
+ \frac{g}{2} \sum_i \left( a_i^2 e^{-i 2\phi_i} + {a^\dagger_i}^2 e^{i 2\phi_i} \right) + H_{\rm C} .
\end{equation}
The effect of the site-dependent phase, namely, breaking time-reversal symmetry, is more apparent after we carry out the gauge transformation
$a_i \to a_i e^{i \phi_i}$, which leads to our final Hamiltonian,
\begin{eqnarray}\label{model}
H &=& \Delta \sum_i a_i^\dagger a_i + \frac{g}{2} \sum_i \left( a_i^2 + {a_i^\dagger}^2 \right) \label{eq.model}
\nonumber \\
& &+ 
\sum_{\substack{i,j \\ (i \neq j)}}
\frac{J_{\rm c}}{|i-j|^3}  a_i^\dagger a_j e^{-i\Delta \phi(j-i)}  .
\end{eqnarray}
Where we have assumed a linear gradient of the phase, such that
\begin{equation}
\phi_i - \phi_j = \Delta \phi (j - i).
\label{eq.gradient}
\end{equation}
Notice that the phase $\Delta \phi$ in the phonon hopping terms cannot be gauged away, due to the presence of the parametric interactions in Eq. \eqref{model}.

Frequency scales for $J_{\rm c}$ depend on the distance between ions \cite{Haze12pra, kiefer2019floquet}. 
In this article, we consider the range $J_{\rm c} = $0.1-10 ($2 \pi$) kHz regime.
As we shall see in the following sections,
the most interesting physics occurs if $\Delta$ and $g$ take similar values as $J_{\rm c}$
Note that this is fully compatible with the assumption of a weak parametric drive, 
$g \ll \omega_{\rm t}$, since typical trapping frequencies are in the 
1-10 (2$\pi$) MHz regime. From this point forward in the manuscript, all simulations will be conducted in units where $J_c=1$. Experimental parameters will only be provided in the sensing section to determine the sensitivities in the appropriate units.

\subsection{Dissipation and Master Equation}
Our model's next ingredient is dissipation in phonon decay, which is required to stabilize the non-equilibrium steady-state. 
Contrary to photonic systems, phonon decay does not naturally occur in ion traps, 
but it can be induced using continuous laser cooling. 
In this process, ions are excited by a laser that is on-resonance with the ions' red-sideband transition, 
while a fast decay re-pumping mechanism re-initializes the system after a phonon has been absorbed. 
This is a continuous version of the experimentally more common stroboscopic scheme, in which ions are subjected to cooling cycles \cite{Leibfried03rmp}. 
We introduce here the laser cooling Liouvillian term and refer to Appendix \hyperref[app:anexoA]{A} for a derivation. 

The following quantum master equation describes the dynamics of the trapped-ion system in the presence of laser cooling 
\begin{equation}\label{eom}
    \frac{d{\rho}}{dt} = -i \left[{H},{\rho}\right]+\mathcal{L}_d({\rho}),
\end{equation}
where the Liouvillian describes phonon decay by laser cooling
\begin{equation}\label{dissipation}
    \mathcal{L}_d({\rho})=\sum_{i}\frac{\gamma}{2}\left(2{a}_i{\rho}{a}^\dagger_i - {a}^\dagger_i{a}_i{\rho} - {\rho}{a}^\dagger_i{a}_i \right),
\end{equation}
with $\gamma$ as the cooling rate. 
The latter can be controlled with the cooling laser's intensity so that $\gamma$ takes values comparable with the other frequency scales in our model (0.1 - 10 (2$\pi$) kHz).
Apart from decay processes such as laser cooling, this model could also account for gain processes, which would be induced by a blue-sideband transition leading to ion heating.

\subsection{External force / electric field}
The last element of our toolbox is a linear term describing an external force \cite{bollinger_dm},
\begin{equation}
H_{\ff}(t) 
= \sum_i F_i(t) \sin(\omega_\ff  t +  \psi_i) \ x_i,
\label{Hf.labframe}
\end{equation}
where we assume that the force is close to resonance with the parametric drive frequency, $\omega_{\rm f} \approx \omega_{\rm d}$, and the
function $F_j(t)$ describes a slow varying amplitude $|\dot{F}_i(t) / F_i(t)| \ll \omega_{\rm t}$. 

We write the position operators $x_j$ in the rotating frame defined in Eq. \eqref{rot.frame} and neglect fast rotating terms so that we get
\begin{equation}
H_{\ff}(t) =  
i \sum_i  \left( \epsilon^*_i(t) a_i  - \epsilon_i(t) a^\dagger_i \right),
\label{Hf}
\end{equation}
where 
\begin{equation}
\epsilon_i(t) =  - \frac{F_i(t) x_0}{2} e^{-i \delta_{\rm f} t - i \psi_i}, \ \ \delta_{\rm f} = \omega_{\rm f} - \omega_{\rm d}.
\label{eq:epsilon}
\end{equation}
We have defined $\delta_\ff$ as the force frequency relative to $\omega_{\rm d}$, which is the reference point of our rotating frame. 

\begin{figure}[h]
    \centering
    \includegraphics[width=0.5\textwidth]{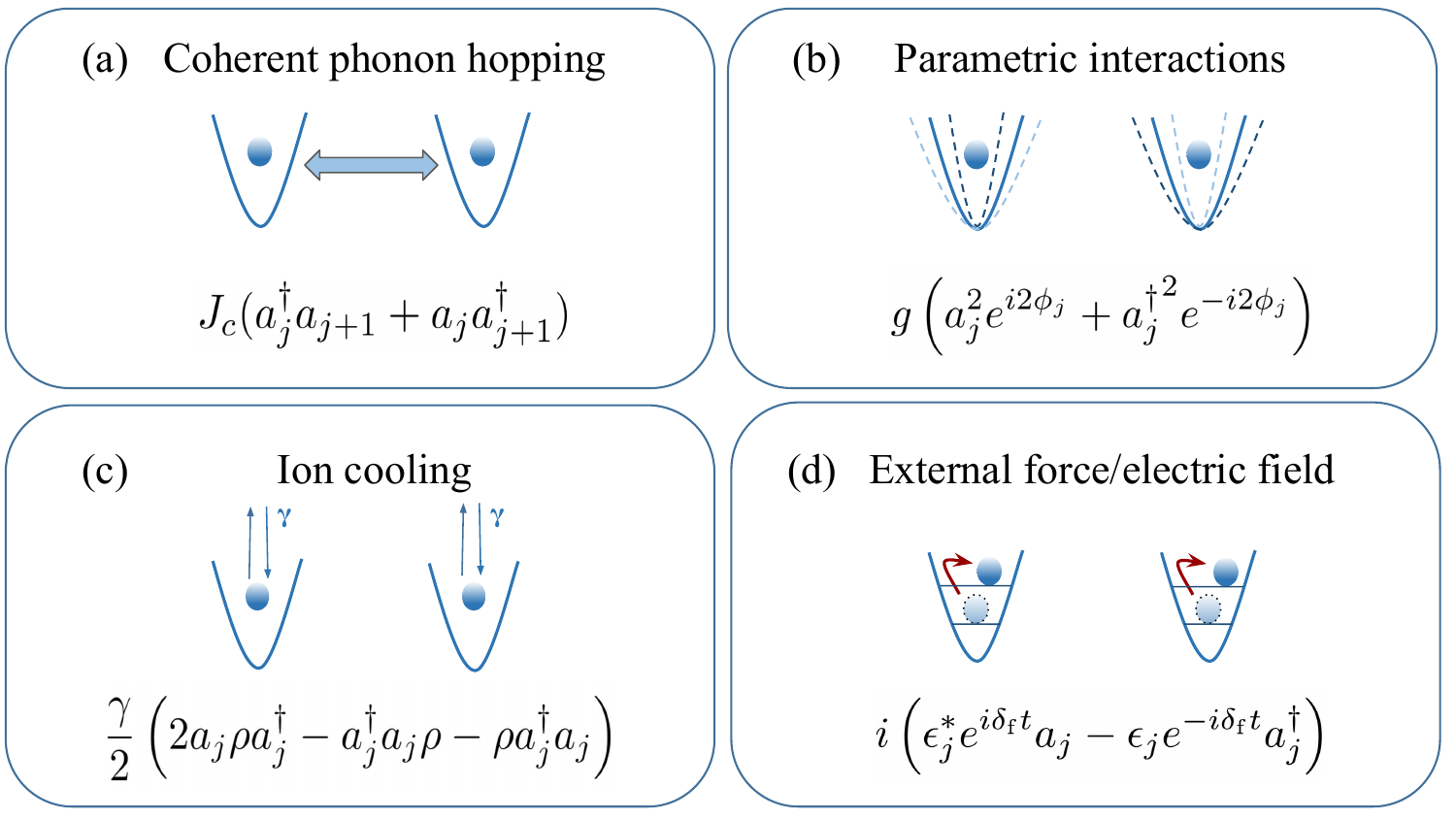}
    \caption{Summary of Hamiltonian and dissipative interactions included in the theoretical model. }
    \label{parm}
\end{figure}

\section{Nonreciprocity and topology in the linear response and steady-state properties}\label{section3}
\begin{figure}[h]
    \centering
    \includegraphics[width=0.5\textwidth]{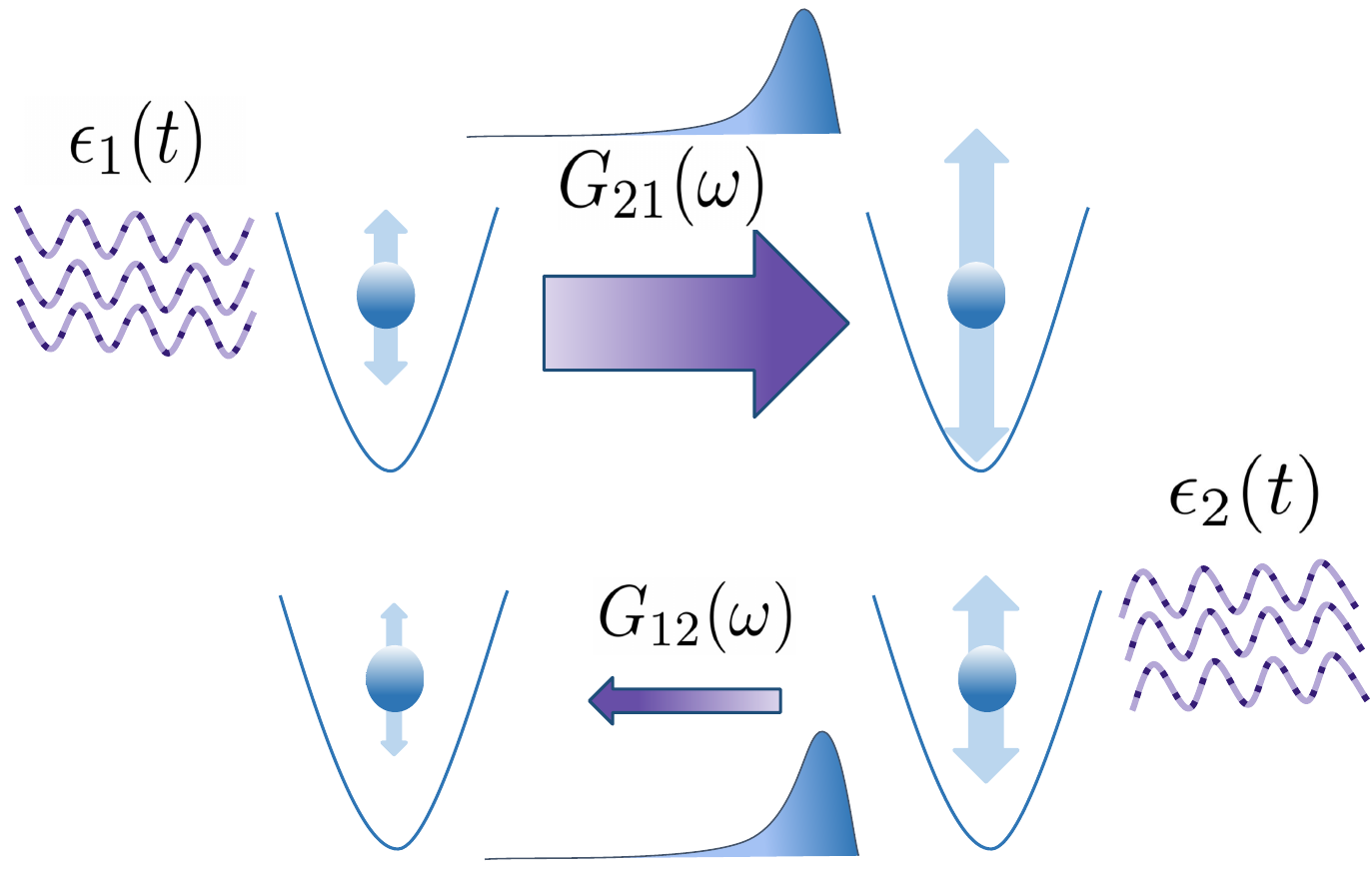}
    \caption{A schematic of the system's response after applying an external field to different ions. The nonreciprocity of the Green's function ($G_{21}\gg G_{12}$) is a hallmark of the topological phases, manifested by the unequal responses when exchanging the force/response sites.}
    \label{epsi}
\end{figure}

The site-dependent phase of the parametric drive in Eq. \eqref{eq:parametric.drive} allows us to break the time-reversal symmetry. 
As we show in this section, this induces a nonreciprocal response of the system (see Fig. \ref{epsi}), as well as nontrivial topological dissipative phases. 
Both nonreciprocity and topology can be detected by:
(i) Studying the response to external perturbations like external forces, which leads to an analogous situation to the transmission of light in photonic systems. Formally, this response function can be described by means of Green's functions, and topological phases can be understood in terms of directional amplification of vibrational excitations.
(ii) Directly measuring properties of phonon observables in the steady-state, such as the phonon number profile along the chain or phononic correlations.

\subsection{Linear response to external forces}
Our setup is an array of coupled vibrational parametric amplifiers of external forces such as the term in Eq. \eqref{Hf}.
If such external force is present, trapped ions get displaced leading to a non-zero coherence in the vibrational modes,
$\langle a_j \rangle \neq 0$. 
We write down the equations of motion for the phononic coherences, 
\begin{align}
\frac{d\langle{a}_i\rangle}{dt}  & = 
  -i \sum_j  J_{i j} \langle a_j \rangle - (i \Delta + \frac{\gamma}{2}) \langle a_i \rangle   
   -  i  g \langle {a}_i^\dagger \rangle  - \epsilon_i(t), 
   \nonumber \\
\frac{d\langle {a}^\dagger_i\rangle}{dt}  & =
  i \sum_j  J^*_{i j} \langle a_j^\dagger \rangle + (i \Delta - 
   \frac{\gamma}{2} )  \langle a^\dagger_i\rangle 
   +  i g \langle a_i\rangle -\epsilon^*_i(t) . \label{e2}
\end{align}
We have defined vibrational coupling matrices,
\begin{eqnarray}
       J_{ij} &\equiv&  \frac{J_{\rm c}}{|i-j|^3} e^{-i\Delta\phi(j-i)}
       \label{eq:G}. 
\end{eqnarray}
We can rewrite the equations of motion in matrix notation as
\begin{equation}\label{comp2d}
    \frac{d}{dt}
    \begin{pmatrix}\langle {\mathbf a}\rangle\\ \langle\bold{a}^\dagger\rangle\end{pmatrix}=
    - i \ \mathbb{H} \begin{pmatrix}
    \langle\bold{a} \rangle\\ \langle\bold{a}^\dagger\rangle\end{pmatrix}-\begin{pmatrix}\boldsymbol{\epsilon}(t)\\ \boldsymbol{\epsilon}^*(t)
    \end{pmatrix},
\end{equation}
where the bold letters represent $N$-dimensional vectors. 
The non-Hermitian $2N \! \times \!  2N$ matrix $\mathbb{H}$ has the following block-structure
\begin{equation}\label{hnh}
    \mathbb{H} = 
    \begin{pmatrix} 
    J + \Delta \mathds{1} - i \frac{\gamma}{2} \mathds{1} &  g \mathds{1} 
    \\
    -g \mathds{1} & -J^* - \Delta \mathds{1} - i \frac{\gamma}{2} \mathds{1}
    \end{pmatrix},
\end{equation}
where $\mathds{1}$ is the $N \!  \times \! N$ identity matrix. 
$\mathbb{H}$ is typically referred to as dynamical matrix or non-Hermitian Hamiltonian. 
It contains all the information necessary to deduce topological and amplification properties. 
The non-Hermitian matrix $\mathbb{H}$ has a particle-hole symmetry, associated to the Nambu index, associated to the pairs
$a_i$, $a_i^\dagger$,
\begin{equation}\label{phs}
\sigma_x \mathbb{H}^* \sigma_x = - \mathbb{H} .
\end{equation}

The Green's function yields the system's linear response and is related to $\mathbb{H}$ by the relation
\begin{equation}
    {\mathbb G}(\omega) = \frac{1}{\omega \mathds{1} - \mathbb{H}} .
\end{equation}
This is a central quantity in the input-output and Keldysh formalisms \cite{ramos2021topological,gomez2023driven, gomez2022bridging}. From the particle-hole symmetry of $\mathbb{H}$ in Eq. (\ref{phs}) we can derive the following property of the Green's function,
\begin{equation}
\sigma_x \mathbb{G}^*(\omega) \sigma_x = - \mathbb{G}(-\omega).
\label{G.sym}
\end{equation}
This implies the block structure 
\begin{equation}
\mathbb{G}(\omega) = 
\begin{pmatrix}
G(\omega) &&\bar{G}(\omega) \\
\bar{G}'(\omega) && G'(\omega) 
\end{pmatrix} .
\label{eq:blocks}
\end{equation}
with sub-blocks satisfying 
\begin{eqnarray}
G'(\omega) &=& - G^*(-\omega) , \nonumber \\
\bar{G}'(\omega) &=& - \bar{G}^*(-\omega),
\end{eqnarray}
due to the symmetry relation in Eq. \eqref{G.sym}

The role of $\mathbb{G}(\omega)$ can be understood by writing Eq. \eqref{comp2d} in frequency space.
Let us define Fourier components 
\begin{eqnarray}
\tilde{a}_i(\omega) = \frac{1}{2\pi} \int a_i(t) e^{i\omega t} dt , \ \
\tilde{\epsilon}_i(\omega) = \frac{1}{2\pi} \int \epsilon_i(t) e^{i \omega t} dt \nonumber .
\end{eqnarray}
We apply those transformations and get, in the steady-state,
\begin{equation}\label{comp}
    \begin{pmatrix}\langle \tilde{\mathbf{a}}(\omega) \rangle \\ \langle \tilde{\bold{a}}^\dagger(-\omega)\rangle \end{pmatrix}=-i \mathbb{G}(\omega) 
    \begin{pmatrix} \tilde{\boldsymbol{\epsilon}}(\omega) \\ \tilde{\boldsymbol{\epsilon}}^*(-\omega)
    \end{pmatrix}.
\end{equation}
This expression relates the coherent vibrational signal with the frequency components of the coherent drive. 

\begin{figure}[h]
    \centering
    \includegraphics[width=0.5\textwidth]{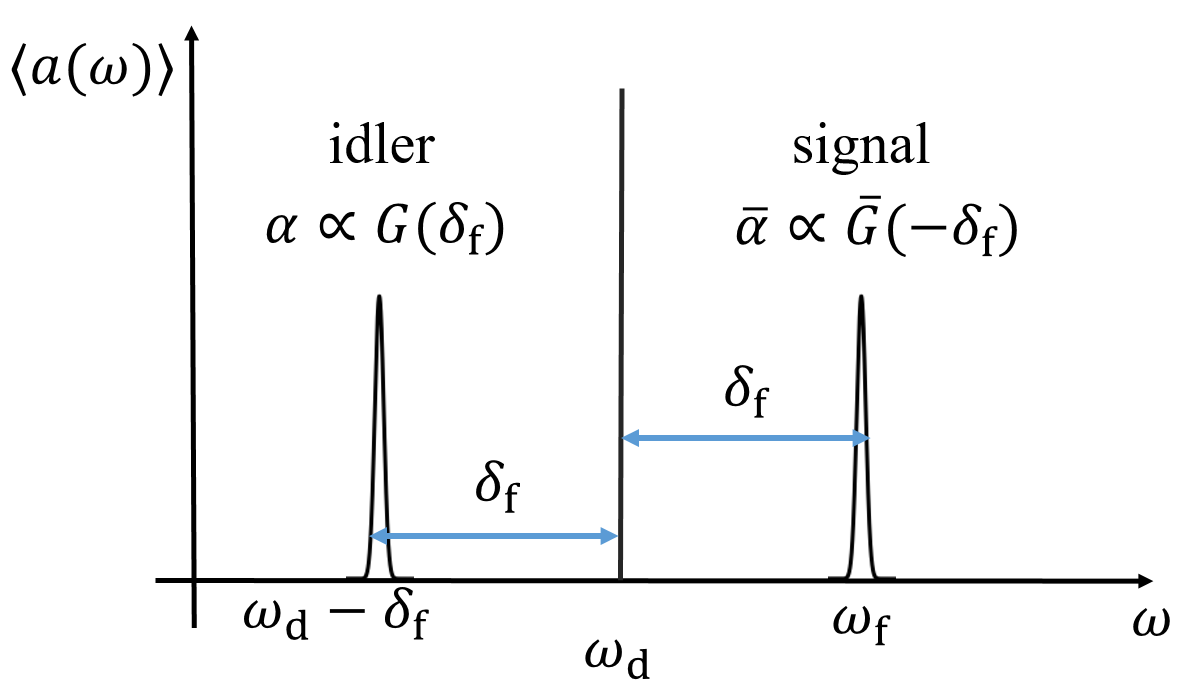}
    \caption{Spectral distribution of the signal and idler contributions to the linear response of the system in terms of the Green's function components.}
    \label{signal-idler}
\end{figure}

Let us use the previous equations to get an explicit result for the steady-state linear response to a monochromatic force. 
According to Eq. \eqref{eq:epsilon}, a monochromatic force leads to a coherent drive that can be expressed, in the rotating frame, as
\begin{equation}
\epsilon_j(t) = \epsilon_j e^{- i \delta_{\rm f} t}.
\nonumber
\end{equation}
The steady-state value of the field operators is a time-dependent oscillation of the form
\begin{equation}
\langle  {\bf a}  (t) \rangle_{\rm ss}  ç
= \boldsymbol{\alpha} e^{- i \delta_{\rm f} t} + \bar{\boldsymbol{\alpha}} e^{i \delta_{\rm f} t},
\end{equation}
with Eq. \eqref{comp} leading to
\begin{equation}\label{comp2}
    \begin{pmatrix} \boldsymbol{\alpha} \\ \bar{\boldsymbol{\alpha}}^* \end{pmatrix}=-i \mathbb{G}(\delta_{\rm f}) 
    \begin{pmatrix} \boldsymbol{\epsilon} \\ 0
    \end{pmatrix}.
\end{equation}
The two components $\boldsymbol{\alpha}$, $\bar{\boldsymbol{\alpha}}$ 
above have the usual signal/idler interpretation
\cite{gardiner_book}, see Figure \ref{signal-idler}. 
By using the sub-block structure of the Green's function, we can actually simplify the response signal to the expression
\begin{equation}
\langle \mathbf{a}(t) \rangle_{\rm ss} = 
- i G(\delta_{\rm f}) e^{- i \delta_{\rm f} t} \boldsymbol{\epsilon} 
- i \bar{G}(- \delta_{\rm f}) e^{i \delta_{\rm f} t} \boldsymbol{\epsilon}^*.
\label{atss}
\end{equation}
The above result is in a frame rotating with $\omega_{\rm d}$. 
We need to go back to the lab frame to obtain the detected signal,
\begin{align}
& \langle \mathbf{a}(t) \rangle_{\rm ss} = 
\\
& - i G(\delta_{\rm f}) e^{- i (\omega_{\rm d} + \delta_{\rm f}) t} \boldsymbol{\epsilon} 
- i \bar{G}(- \delta_{\rm f}) e^{- i (\omega_{\rm d} - \delta_{\rm f}) t} \boldsymbol{\epsilon}^*. \nonumber
\end{align}
From this result, we can explicitly interpret the Green's function blocks $G(\delta_{\rm f})$ and 
$\bar{G}(\delta_{\rm f})$ as the signal and idler amplification terms, respectively.


\subsection{Vibrational correlation functions and steady-state properties}

We focus now on observables such as phonon numbers and phonon correlations, which carry information on the topological phases of the system, even in the absence of any coherent driving. Those quantities will also be relevant to calculating the noise in quantum sensing applications (section \ref{section6}).

In order to simplify the notation we define operators $\mathsf{a}_\mu$ with 
$\mu =  1, \dots, 2 N$ in the Nambu basis, 
\begin{eqnarray}
\mu = j &\to& \mathsf{a}_\mu = a_j , \nonumber \\
\mu = j + N &\to& \mathsf{a}_\mu  = a_j^\dagger. \nonumber
\end{eqnarray} 
We define the 
$2N \! \times \! 2N$ correlation matrix
\begin{equation}\label{pcf}
C_{\mu \nu} \equiv 
\langle \mathsf{a}^\dagger_\mu \mathsf{a}_\nu \rangle - 
\langle \mathsf{a}^\dagger_\mu \rangle 
\langle \mathsf{a}_\nu \rangle.
\end{equation}
Notice that in the absence of external forces, the second term of the right hand side of equation (\ref{pcf}) vanishes. This matrix has a block structure in terms of normal ($N$) and anomalous terms ($M$),
\begin{equation}\label{C}
    C = 
    \begin{pmatrix} 
    N &  M 
    \\
    M^*  & N + \mathds{1}
    \end{pmatrix},
\end{equation}
where $N_{ij} = \langle a^\dagger_i a_j \rangle$ and $M_{ij} = \langle a^\dagger_i a^\dagger_j \rangle$.   

$C_{\mu \nu}$ can be calculated by solving the equations of motion derived from the system's Liouvillian (see Appendix \ref{app:anexoB}). 
However, the Green's function formalism also also allows us to evaluate this quantity. The following expression can be derived using the Keldysh formalism \cite{gomez2022bridging} or by analyzing the master equation (see Appendix \ref{app:anexoB}),
\begin{equation} \label{eq:Cintegral}
C = \int \frac{d\omega}{2 \pi}  
\mathbb{G}^*(\omega)  \begin{pmatrix}
0 && 0 \\
0 && \gamma \mathds{1}
\end{pmatrix}  \mathbb{G}^{\rm T}(\omega) .
\end{equation}
Beyond its practical use, the above equation highlights the interpretation of correlations as determined by the effect of incoherent processes propagated by the Green's function and averaged over all frequencies.

\subsection{Stability}
Finally, we discuss the stability criterion under the parametric driving considered in this work. 
In bosonic systems, the underlying Fock space is infinite-dimensional, such that the previous expectation values are formally unbounded \cite{ughrelidze2024interplay}. 
Under conditions of strong parametric driving, the amplification of excitations can lead to unstable phases with ill-defined steady-state of our dissipative system. 
The stability criterion can be obtained from the Lyapunov equation provided in 
Appendix \hyperref[app:anexoB]{B}, 
\begin{equation}
\Im{\lambda_\mu} < 0, \ \  \forall \mu ,
\end{equation}
where $\lambda_\mu$ are the eigenvalues of $\mathbb{H}$. 
In general, the system is stable when there are few ions in the chain or the dissipation is relevant enough. 

\section{Dissipative topological phases in trapped ion chains}
\label{section5}
\subsection{Topological amplification theory}
The non-equilibrium phases of the parametric trapped ion chain can be understood through the lens of the topological amplification theory previously introduced in \cite{porras2019topological, gomez2023driven}. 
The starting point of this formalism is the singular value decomposition (SVD) of the inverse of the Green's function,
\begin{equation}\label{svd}
     \mathbb{G}^{-1}(\omega) = \omega \mathds{1} - \mathbb{H} = U S V^\dagger.
\end{equation}
$U$ and $V^\dagger$ are unitary matrices and $S$ is a positive diagonal matrix, $S_{nm} = s_{n} \delta_{nm}$. 
The SVD is a convenient basis for expressing the Green's function,
\begin{equation}
\mathbb{G}(\omega) =
V S^{-1} U^\dagger .
\label{svd}
\end{equation}
As proved in \cite{porras2019topological,gomez2023driven}, the properties of $\mathbb{G}(\omega)$ are determined by the following topological invariant (winding number)
\begin{equation}\label{winding1}
    \nu(\omega)
    = \Im \int_{-\pi}^{\pi}
    \frac{dk}{2\pi}
    \Tr \partial_k \log \left( \omega \mathds{1} - \mathbb{H}(k) \right),
\end{equation} 
where $\mathbb{H}(k)$ is the non-Hermitian dynamical matrix with periodic boundary conditions in the plane-wave basis. 
This topological invariant has been introduced in the context of non-Hermitian point-gap topology \cite{Sato2023review} as well as in the description of topological amplification \cite{gomez2022bridging,gomez2023driven}. 
Nontrivial values $\nu(\omega) \neq 0$ imply the appearance of quasi zero-singular values and edge-singular vectors, 
which are the dissipative equivalent to the zero-energy modes of topological insulators \cite{hasan2010colloquium}.
This result leads to a bulk-boundary correspondence in non-Hermitian systems, although for the SVD rather than the eigensystem (see Appendix \ref{app:anexoE} for a derivation of this result).

\begin{figure}[h]
    \centering
    \includegraphics[width=0.5\textwidth]{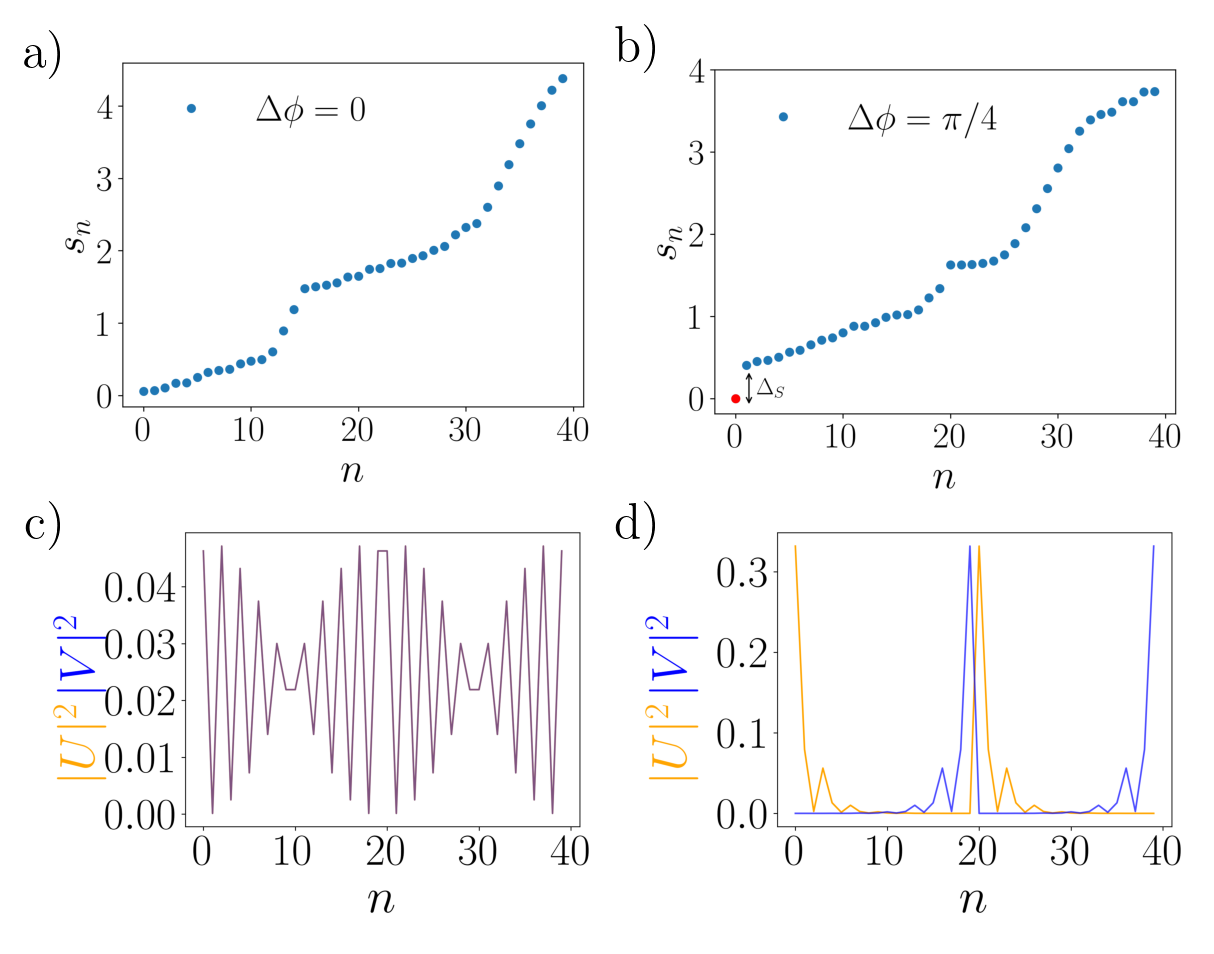}
    \caption{Singular value decomposition of $\mathbb{H}$ for $J_{\rm c}=\Delta=g=\gamma=1$ and  $\omega=0$ obtained from numerical simulations of an $N=20$ ion chain. (a) Trivial configuration ($\Delta \phi=0$) with no zero-singular values. (b) Topological configuration ($\Delta \phi=\pi/4$) with a zero-singular value (red dot) separated from the bulk by a gap. (c) Delocalization of a singular vector from the trivial configuration ($\Delta \phi=0$). (d) Localization of the edge state showing topological properties ($\Delta \phi=\pi/4$). }
    \label{prlr}
\end{figure}

Zero-singular values associated to the edge modes, $s_n : n\in {\cal N}_{\rm e}$, decrease exponentially with the system size, $s_{n_{\rm e}} \propto e^{- N/\xi}$, with $\xi$ the localization length of the edge-singular state
Furthermore, they are separated by a singular value gap from the rest of the singular value spectrum. 
\cite{porras2019topological,gomez2023driven}.
In Fig. \ref{prlr}, we plot an example of topologically trivial (a) and nontrivial (b) singular values. 
The case $\Delta \phi = \pi/4$ in Fig. \ref{prlr} (b) shows the appearance of a zero-singular value. In Fig. \ref{prlr} (c, d), we present the singular vectors and show that zero-singular vectors correspond to localized states in the topological case.

\begin{figure}[h]
    \centering
        \includegraphics[width=0.5\textwidth]{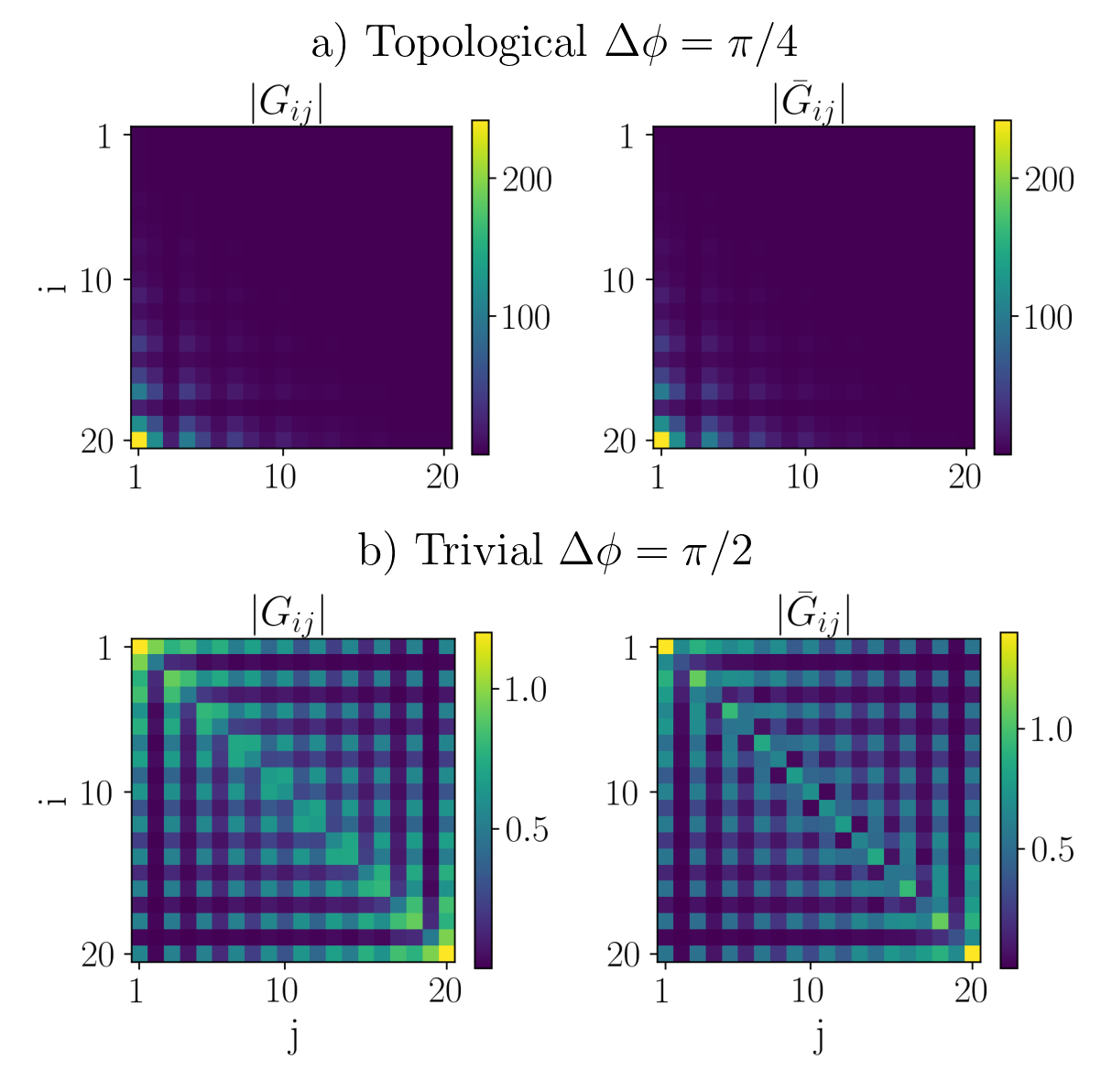}
    \caption{Module of the first and second quadrants of the Green's function matrix elements for different values of $\Delta \phi$, and $\Delta = J_{\rm c} = g = \gamma = 1$, $\omega = 0$, and $N = 20$. 
    (a) The topological regime ($\Delta \phi=\pi/4$) shows strong nonreciprocal amplification of $|G_{ij}|$ and $|\bar{G}_{ij}|$. (b) The trivial case ($\Delta \phi=\pi/2$) shows a reciprocal response of the Green's function.}
    \label{colormapGf}
\end{figure}

In the topological phase, zero-singular values and their corresponding edge-singular vectors dominate the sum over $n$ in Eq. \eqref{svd}. 
In the case where there is a single zero-singular value, $n_e$, the Green's function can be approximated by
\begin{equation}
\mathbb{G}_{\mu \nu}(\omega) \approx V_{\mu n_{\rm e}} s_{n_{\rm e}}^{-1} U^*_{\nu n_{\rm e}}.
\label{svd3}
\end{equation}
Eq. \eqref{svd3} is the mathematical formulation of the phenomenon known as topological amplification. To understand this effect, we first notice that the edge-singular vectors $U_{j n_{\rm e}}$, $V_{i n_{\rm e}}$ are spatially localized at opposite edges of the chain, see Fig. \ref{prlr} for an example. 
This is a direct consequence of topological insulator theory, as shown in our previous works \cite{porras2019topological,ramos2021topological}. 
The localization of edge-singular vectors leads to the Green's function being highly nonreciprocal and directional. 
To understand more explicitly how such directionality arises in the topological amplification regime, let us focus on the properties of $G(\omega)$, 
the upper-left block in Eq. \eqref{eq:blocks}, and write the spatial dependence of the singular vectors by approximating them by exponential functions localized at both edges of the chain,
$V_{i n_{\rm e}} \propto e^{-(N-i)/\xi}$, 
$U_{j n_{\rm e}} \propto e^{-j/\xi}$. Although finite-size corrections modify these expressions, they allow us to get an approximate understanding of the structure of the Green's function,
\begin{equation}
G_{i j}(\omega) \propto e^{-(N-i)/\xi} e^{N/\xi} e^{-j/\xi}.
\label{svd2}
\end{equation}
$G_{N1}\approx e^{(N-1)/\xi}$ is exponentially enhanced compared to $G_{1N} \approx e^{-(N-1)/\xi}$, which implies a strong directional, nonreciprocal, response. 
In Fig. \ref{colormapGf}, we corroborate this result by presenting a numerical calculation of matrix elements of the Green's function in both topologically trivial and nontrivial cases.
Fig. \ref{colormapGf}(a) shows strong localization of the Green's function in the topological phase for values $i \approx N$, $j \approx 1$. In contrast, in Fig. \ref{colormapGf}(b), we observe a fully reciprocal behavior of the Green's function in a topologically trivial case. 

\begin{figure}[h]
    \centering
        \includegraphics[width=0.5\textwidth]{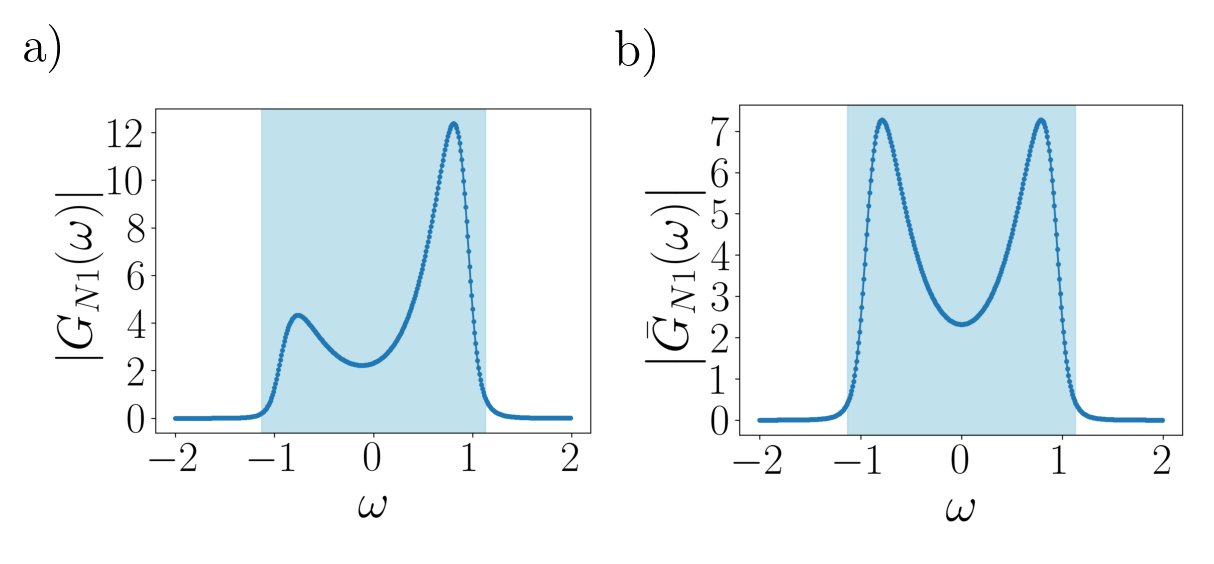}
    \caption{Module of $G$ and $\bar{G}$ as a function of $\omega$. The blue-shaded region corresponds to a topological regime with $\nu(\omega) \neq 0$, where directional amplification properties are guaranteed. The remaining parameters are $J_{\rm c}=\Delta=g=1$,$\gamma=1.8$, $\Delta \phi=\pi/4$, and $N=20$.}
    \label{Gw}
\end{figure}

Topological amplification depends on the frequency at which the system is probed, as explicitly shown by the winding number in Eq. \eqref{winding1}.
Frequency intervals for which $\nu(\omega) \neq 0$ are regions at which the Green's function shows topological amplification. In Fig. \ref{Gw}, we present an example of the dependence of the Green's function on the frequency and show that the system's response is enhanced at specific values $\omega \neq 0$. 
This will have substantial consequences when calculating the quantum sensing capabilities of the parametric trapped ion chain in section \ref{section6}.

\subsection{Phase diagram of the trapped ion parametric chain}
Let us investigate the topological phase diagram under the lens of the topological amplification theory discussed above. 
We consider values $J_{\rm c} = g = \Delta = 1$ and explore the phase diagram as a function of $\Delta \phi$ and $\gamma$, limiting our study to the resonant case 
$\omega = 0$. 
We have numerically found that this parameter choice is especially convenient to obtain topological phases. 
In particular, our calculations show that condition $\Delta \neq 0$ has to be fulfilled to obtain non-trivial values of $\nu(\omega)$

\begin{figure}[h]
    \centering
    \includegraphics[width=0.5\textwidth]{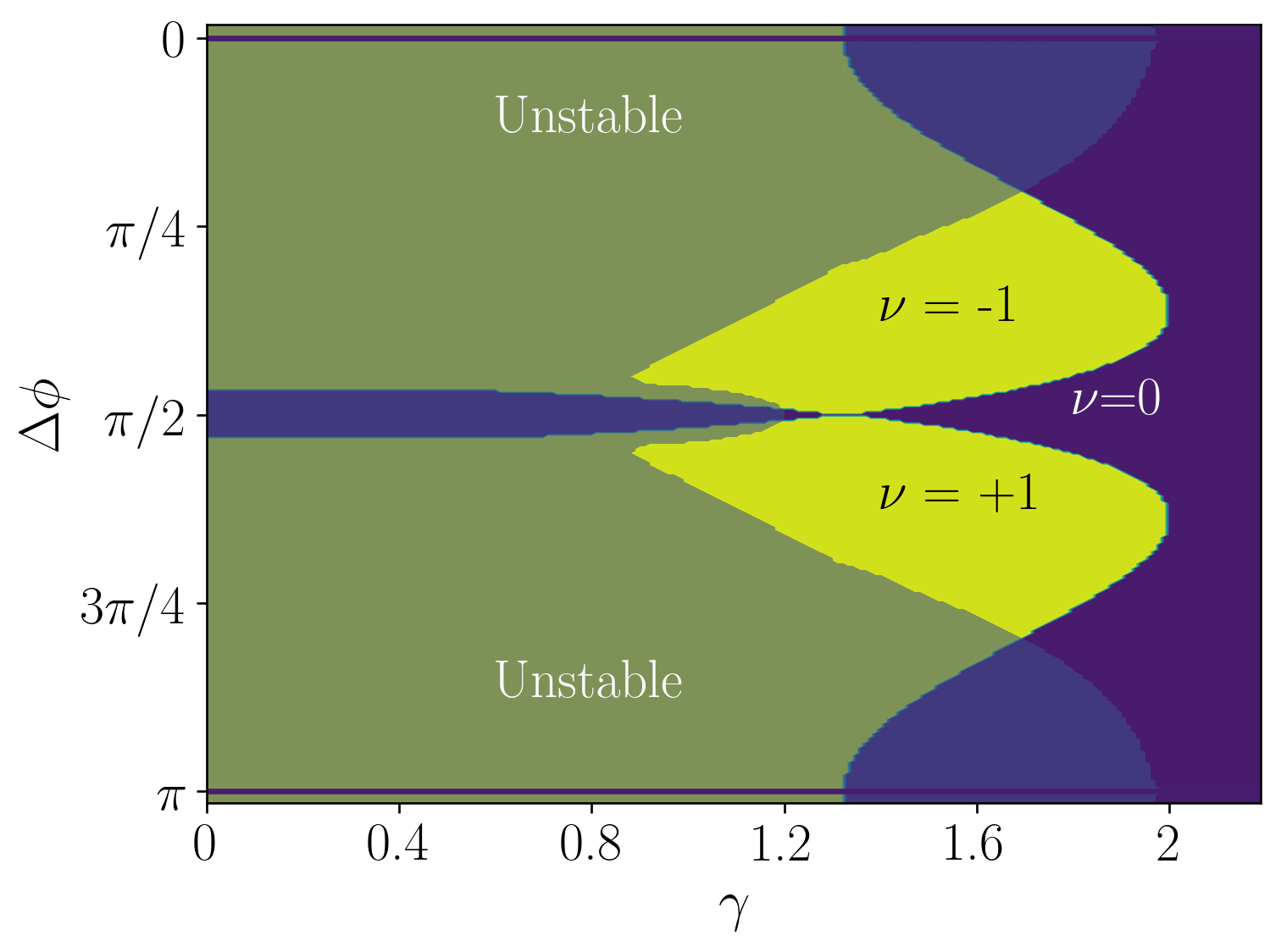}
    \caption{Topological phase diagram in the $\gamma$-$\Delta \phi$ parameter space with $J_{\rm c} = \Delta = g = 1$ and $\omega=0$, calculated in a $N = 20$ ions chain. 
    It presents two nontrivial topological phases with winding numbers $\nu=\pm 1$ (yellow), a topological unstable phase (green), a trivial stable phase (dark blue), and a trivial unstable phase (light blue)  $N=20$.}
    \label{gammaphidd}
\end{figure}

In Fig. \ref{gammaphidd}, we plot topological phases within the parameter space spanned by $\gamma$ and $\Delta \phi$. We calculate the winding number
and find two topological nontrivial phases characterized by $\nu(0) = \pm 1$  (in yellow-green). We find areas (yellow) where the system is stable within those topological regions.

\begin{figure}[h]
    \centering
        \includegraphics[width=0.45\textwidth]{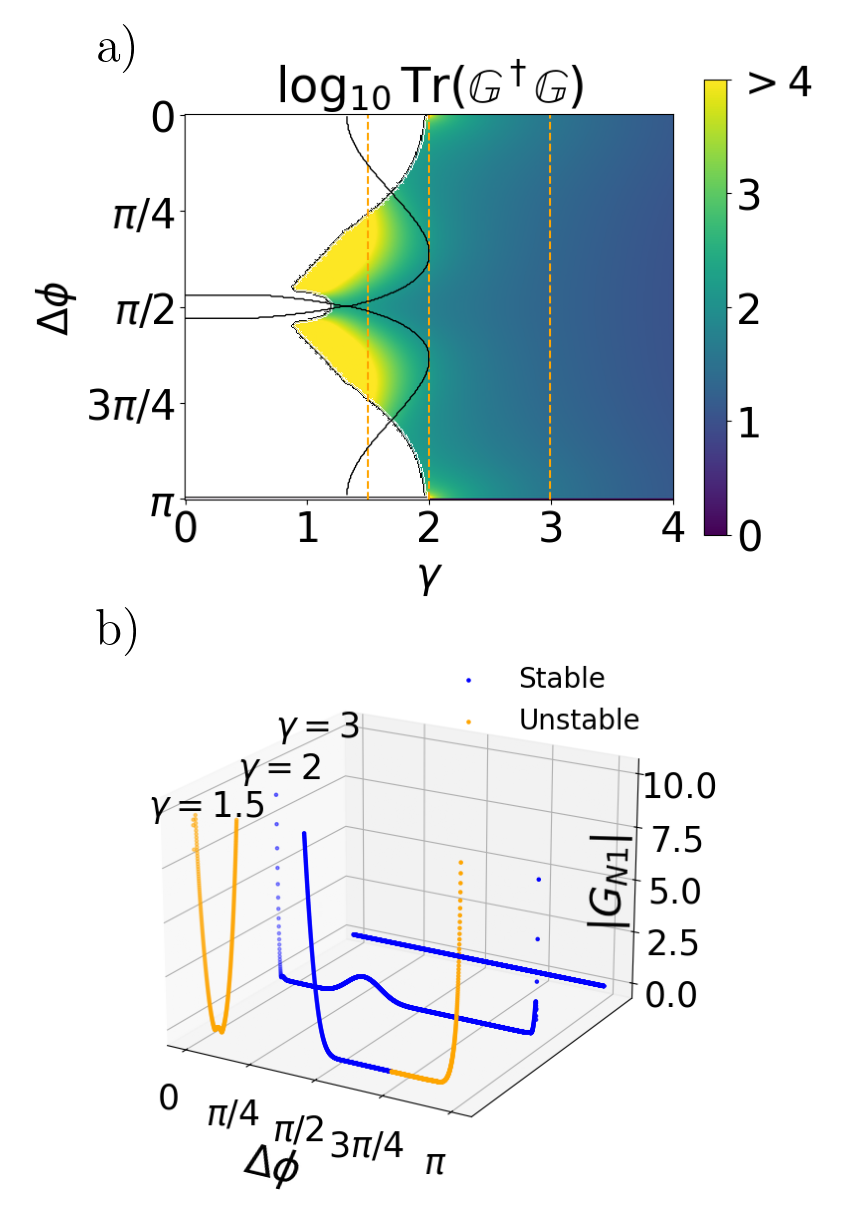}
    \caption{(a) Squared Frobenius norm of the Green's function in logarithmic scale on the $\gamma$ - $\Delta \phi$ phase space. The solid black contour delineates the topological regime obtained from Fig. \ref{gammaphidd}. (b) Fixing certain values of $\gamma$, we focus on the dependence of $|G_{N1}|$ on $\Delta \phi$, showing nonreciprocity for values inside the topological regime. The remaining parameters are $J_{\rm c} = \Delta = g = 1$, $\omega = 0$, and $N=20$. The orange points are unstable, thus the steady state is ill-defined.}
    \label{recre}
\end{figure}

To quantify the amplification effect in topologically stable regions, we calculate the squared Frobenius norm of the Green's function, shown in Fig. \ref{recre}(a) and show that it takes large values at topological regions following the predictions from topological amplification theory. 
In Fig. \ref{recre}(b), we show results for the Green's function matrix element $G_{N1}(0)$, showing the end-to-end amplification effect.

\begin{figure}[h]
    \centering
        \includegraphics[width=0.45\textwidth]{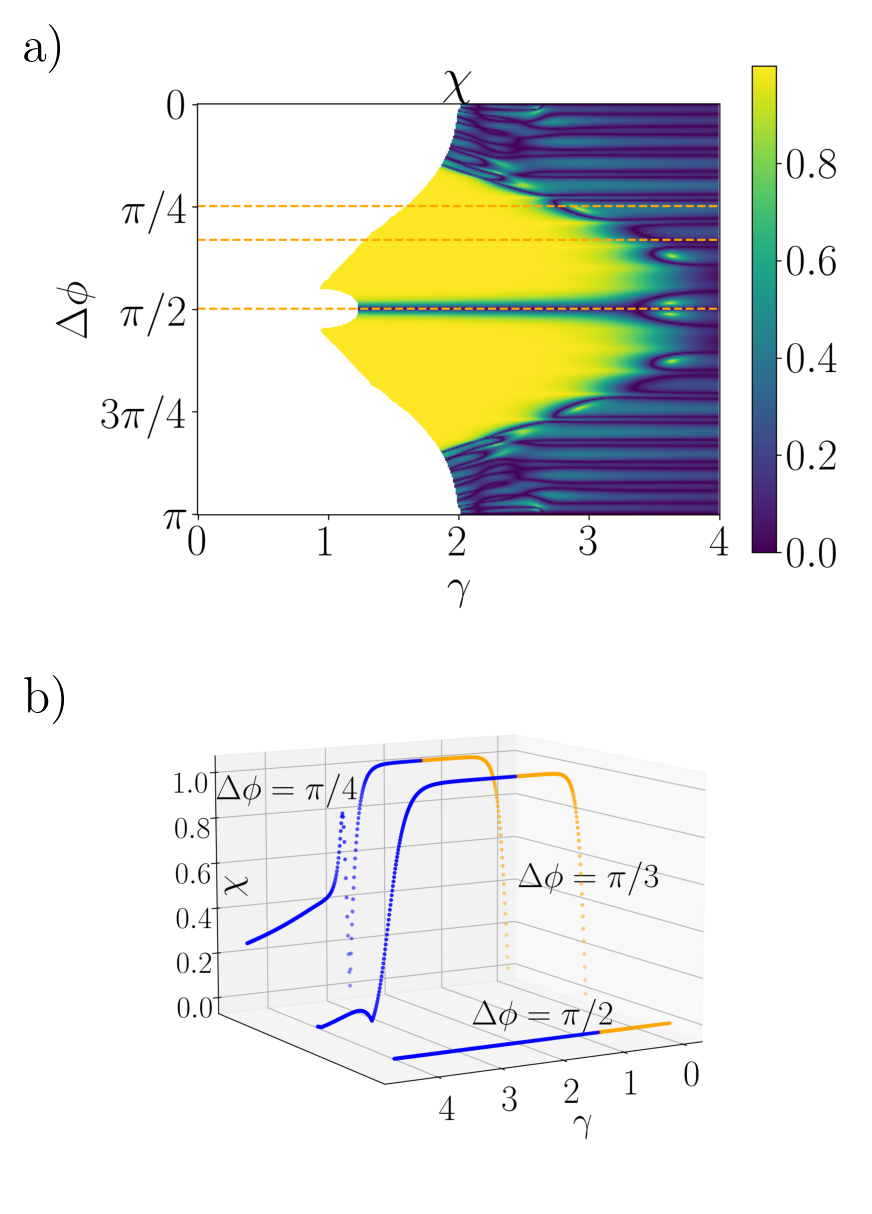}
        \caption{(a) Nonreciprocity phase diagram. In certain regions of the $\gamma$ - $\Delta \phi$ space, the system is nonreciprocal since $\chi > 0$. (b) Fixing certain values of $\Delta \phi$, we focus on the dependence of nonreciprocity on the dissipation coefficient $\gamma$. The remaining parameters are $J_{\rm c}=\Delta=g=1$, $\omega=0$, and $N=20$. The orange points are unstable, thus the steady state is ill-defined.}
    \label{rere}
\end{figure}

Another feature enhanced by topology is the nonreciprocity of the Green's functions which accounts for the directionality of the vibrational signal. 
We quantify the nonreciprocity of the Green's function between ions $i=1$ and $i=N$ using the coefficient
\begin{equation}
\chi = \frac{||G_{N1}|-|G_{1N}||}{||G_{N1}|+|G_{1N}||}.
\end{equation}
In Fig. \ref{rere}, we show that the system's nonreciprocal response is also enhanced close to the topological regions. 
However, nonreciprocity seems to be a feature that survives in the trivial regions since it is related to breaking time-reversal symmetry, even though the latter may not necessarily imply non-trivial values of $\nu(\omega)$. 

\subsection{Steady-state phonon number observables}
\begin{figure}[h]
    \centering
    \includegraphics[width=0.5\textwidth]{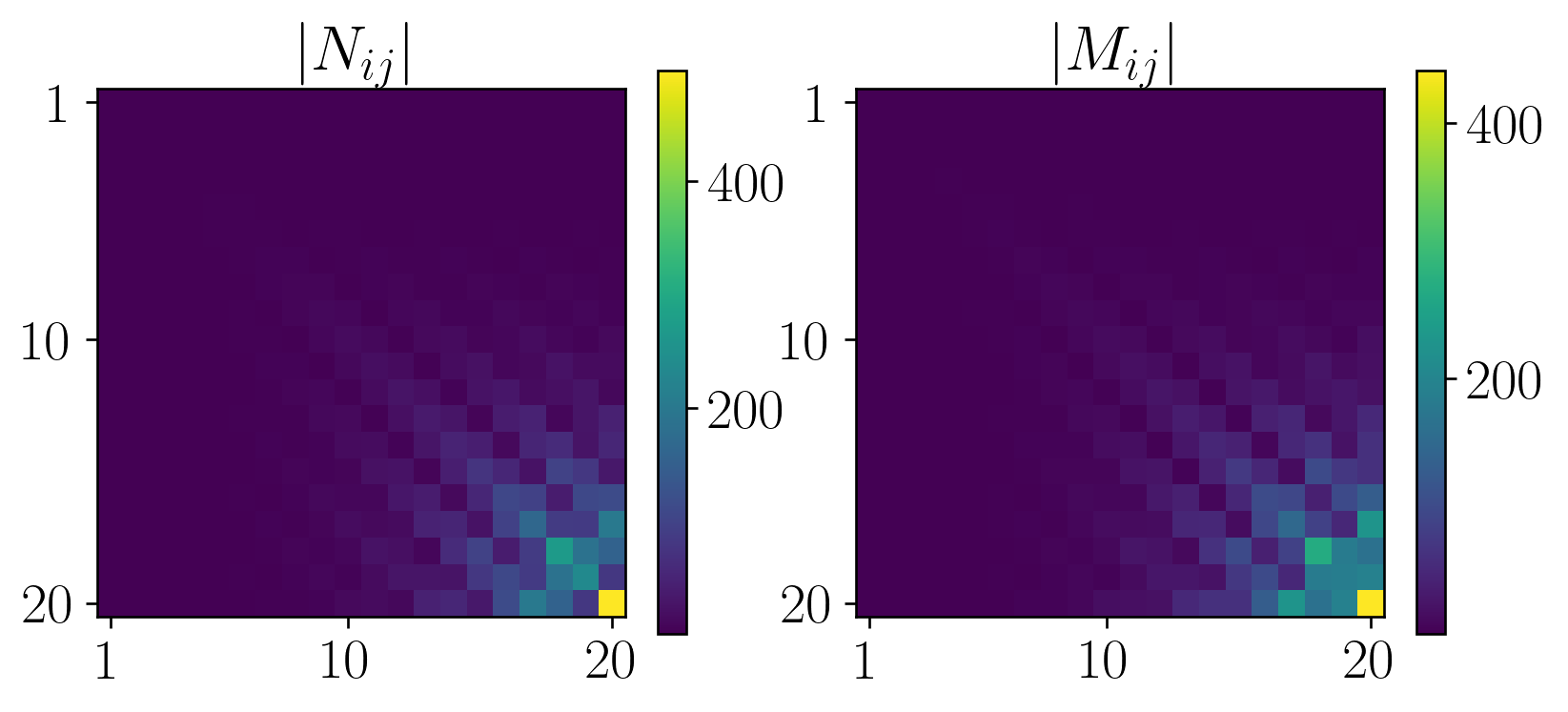}
    \caption{Correlation matrix $C$ blocks for the set of parameters given by the parameters $J_{\rm c}=\Delta=g=1$, $\gamma=1.7$,  $\Delta \phi=\pi/4$ and $N=20$. It shows the topological directional amplification of the submatrices $N_{ij}=\langle a^\dagger_i a_j \rangle$ and $M_{ij}=\langle a^\dagger_i a^\dagger_j \rangle$ in the steady-state.}
    \label{pcompo}
\end{figure}

We now consider the steady-state phonon correlation matrix $C$ from Eq. \eqref{pcf}.
It can be expressed as a convolution of Green's functions at different frequencies $\omega$, see Eq. \eqref{eq:Cintegral}.
In topologically non-trivial phases, $C$ will reproduce the asymmetrical spatial structure that arises due to the non-reciprocity of the Green's function. 
In  Fig. \ref{pcompo}, we show a numerical calculation of the diagonal and non-diagonal blocks of $C$ in a topological regime. Both components are enhanced close to the edge of the chain towards which topological amplification occurs. 

\begin{figure}[h]
    \centering
    \includegraphics[width=0.5\textwidth]{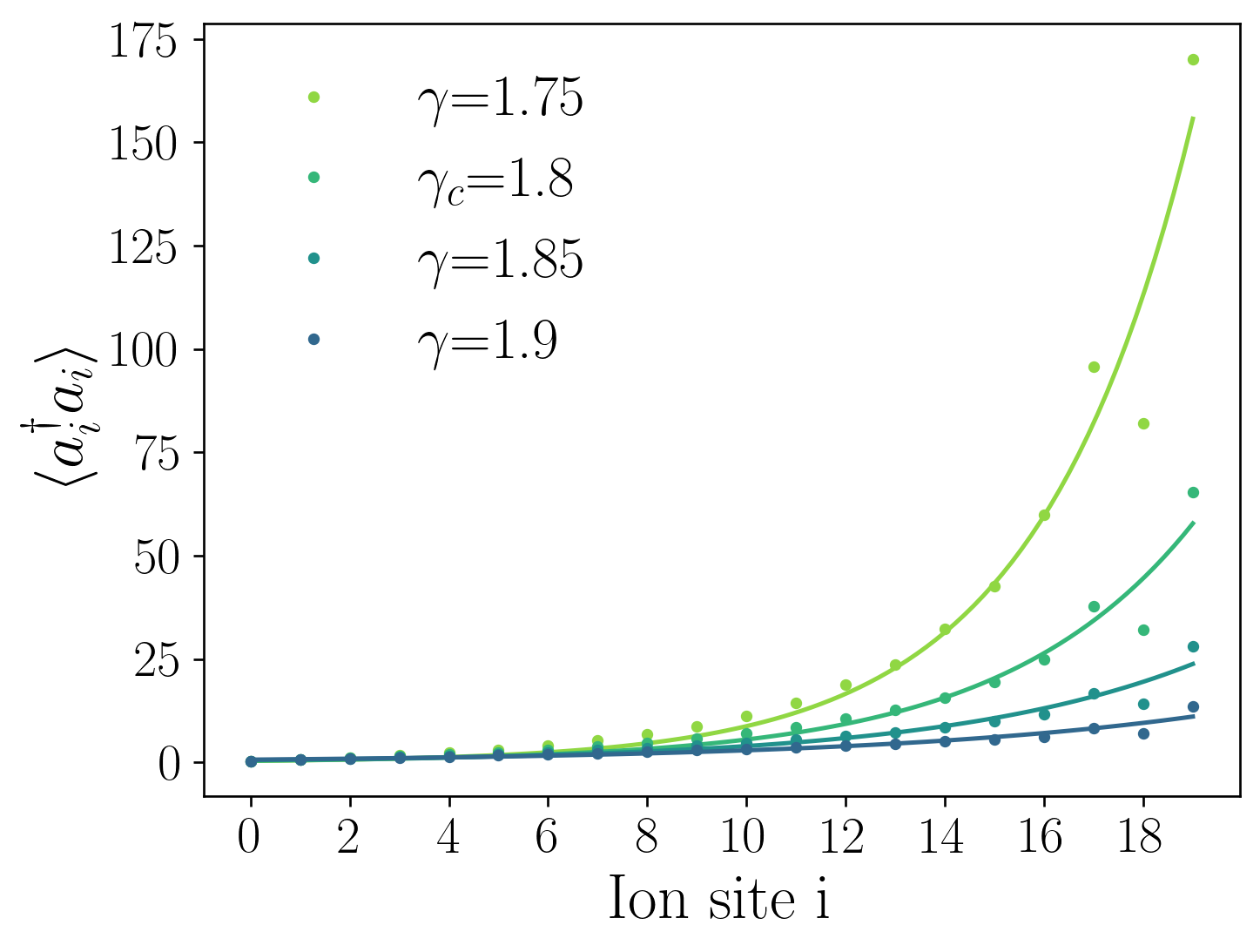}
    \caption{Phonon number average as a function of the array site for different values of the local losses $\gamma$. The fixed parameters are $\omega=0$, $\Delta=g=J_{\rm c}=1$ , $\Delta \phi=\pi/4$ and $N=20$. The solid lines represent an exponential fit to the simulated data (points). }
    \label{pna}
\end{figure}

To observe this effect more explicitly, let us focus on the diagonal terms of the first quadrant of the correlation matrix, which corresponds to the phonon number at each site, $N_{jj} = \langle a^\dagger _j a_j \rangle$. 
In Fig. \ref{pna}, we plot the spatial dependence of $N_{jj}$ for values around the critical value $\gamma_c \approx 1.821$, separating the non-topological ($\gamma >  \gamma_c$ ) and topological phases ($\gamma <  \gamma_c$ ). The plot shows the exponential enhancement of the phonon number as we cross the phase transition and enter the topological amplification regime. 

\begin{figure}[h]
    \centering
    \includegraphics[width=0.5\textwidth]{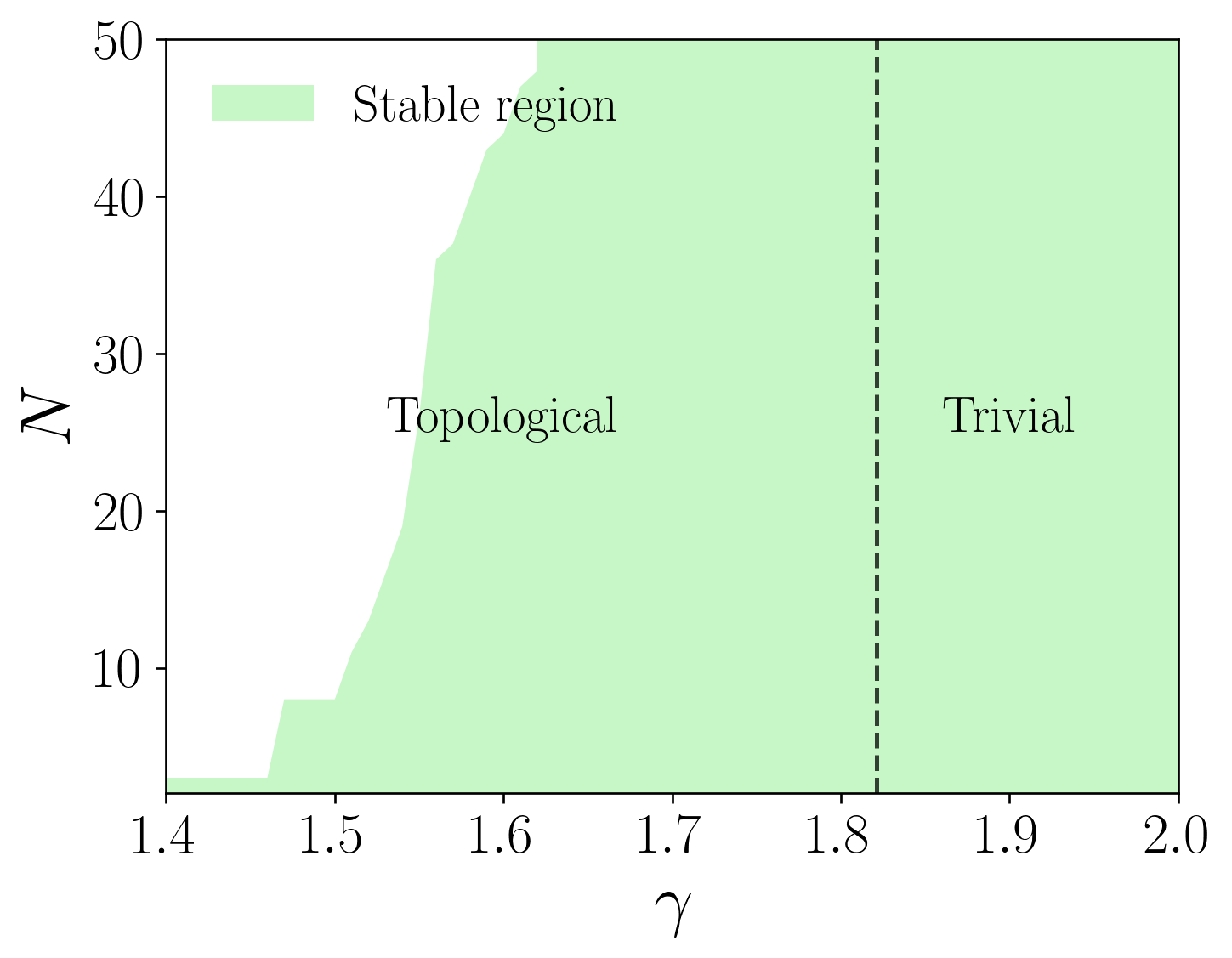}
    \caption{Stability diagram as a function of the local dissipation $\gamma$ and the system size $N$. The remaining parameters are $J_{\rm c}=\Delta=g=1$,  $\omega=0$ and $\Delta \phi=\pi/4$. }
    \label{Nstability}
\end{figure}

Finally, we address the system's stability as a function of system size. In Fig. \ref{Nstability}, we show that the overlap between the topological and stable regions decreases as the system size is enlarged. However, within the range of chain lengths considered in this work (up to $N = 50$), there is an interval of 
$1.538 > \gamma > 1.821$ where the topological amplification regime is stable.  


\section{Quantum sensing}\label{section6}
\begin{figure}[h]
    \centering
    \includegraphics[width=0.5\textwidth]{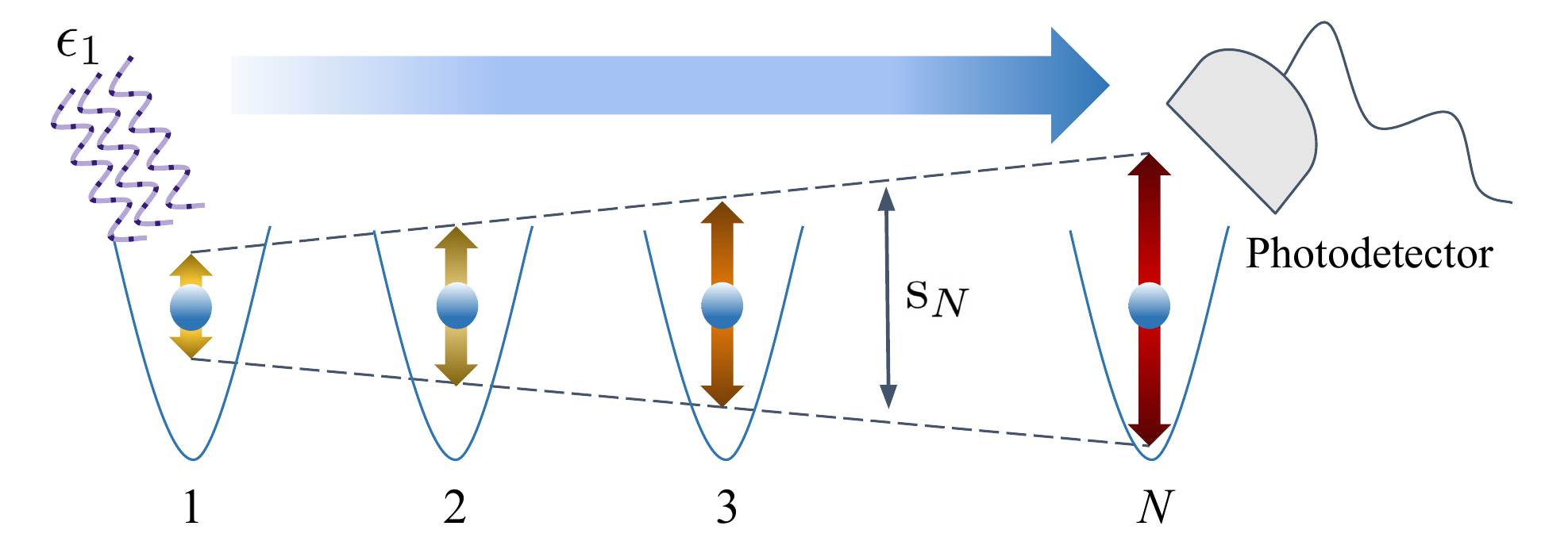}
    \caption{Schematic of the topological sensing protocol. We illuminate the first $^{25}$Mg$^+$ ion with an external field $\epsilon_1$ and measure the time-averaged amplitude of the $N$'th ion's displacement, $\mathrm{s}_N$, through the fluorescence emitted.}
    \label{schemeard}
\end{figure}

The dissipative topological phases of the parametric ion chain studied in the previous section can be applied in the measurement of ultra-weak forces \cite{ivanov2016high, di2023critical, bollinger_dm}.
Most previous works measure the ion's displacement by using a spin-phonon coupling that maps displacements into the spin degree of freedom, which photoluminescence techniques can measure.  
Here, we benefit from the strong directional amplification of vibrations to enhance the ion displacement so that it can be detected directly by fluorescence measurements (see the general setup in Fig. \ref{schemeard}).
We consider a scheme in which the first $^{25}$Mg$^+$ ion of the chain ($i = 1$) acts as a detector of an ultra-weak force, and the signal propagates up to the final ion ($i = N$) that is measured. Variations of this approach could also be considered, such as amplifying the signal by using more ions as detectors.

Since our scheme relies on measuring the position of the $N$'th ion, let us see how our formalism applies to predicting that observable. The expectation value of the last ion's position operator is
\begin{equation}
\langle x_N(t) \rangle = 2 x_0 {\Re} \left( \langle a_N(t) \rangle \right).
\label{eq:position}
\end{equation}
We assume that the driving force has a detuning $\delta_\ff$ relative to the parametric driving, $\omega_{\rm d}$. 
We work in the lab frame in which position measurements take place. 
In terms of our Green's function formalism, the phonon coherence in 
Eq. \eqref{eq:position} can be written like
\begin{eqnarray}
\langle a_N(t) \rangle= &  \label{aNt} \\ 
= - i & \! \left( G_{N1}(\delta_\ff) 
e^{-i (\omega_{\rm d} + \delta_\ff) t} + \bar{G}_{N1}(-\delta_\ff) e^{- i (\omega_{\rm d} - \delta_\ff) t}\right) \epsilon_1. \nonumber 
\end{eqnarray}
The amplitude of the displacement will ultimately depend on both the amplitude and frequency of the ultra-weak force, as well as the trapped-ion parametric chain parameters. 

In the lab frame, $\langle x_N(t) \rangle$ oscillates with the parametric drive frequency, $\omega_{\rm d}$, 
so resolving the ion's oscillation would involve fluorescence measurements resolved with $\mu$s time resolution, which would be experimentally challenging. 
Instead, we will assume that the fluorescence is measured during a long integration time, $T \gg \omega_{\rm d}^{-1}$, such that the integrated fluorescence measures the amplitude of the detector ion's oscillation, $\mathrm{s}_N$, which can be estimated as
\begin{eqnarray}
\mathrm{s}_N^2(\delta_\ff)  &=& \frac{1}{T} \int_0^T \langle x_N(t') \rangle^2 dt' .
\label{sigmasN}
\end{eqnarray}
The integration can be calculated using Eq. \eqref{aNt}. In the resonant case ($\delta_\ff = 0$) we get
\begin{equation}
\mathrm{s}_N^2(0)  = \frac{(2 x_0\epsilon)^2}{2}(|G_{N1}(0) + \bar{G}_{N1}(0)|^2) ,
\label{sigmaxN.resonant}
\end{equation}
whereas in the non-resonant case ($\delta_\ff \neq 0$), and assuming long integration times such that $T \gg \delta_{\ff}^{-1}$, we get
\begin{equation}
\mathrm{s}_N^2(\delta_\ff) = \frac{(2 x_0\epsilon)^2}{2}(|G_{N1}(\delta_\ff)|^2+|\bar{G}_{N1}(-\delta_\ff)|^2).
\label{sigmaxN.nonresonant}
\end{equation}
In the intermediate case with $\delta_\ff \neq 0$, but $1/T > \delta_\ff$, an expression dependent on $T$ would be obtained. This regime would be more challenging to analyze, and, in any case, we will find below that the long-integration time limit is well justified with typical experimental parameters.

\begin{table*}[t]
\centering
\begin{tabular}{|c|c|c|c|c|c|c|}
\hline
\textbf{$N$}& \textbf{Energy scale [kHz]} & \textbf{$\tau$ [s]}& \textbf{$F^{\rm q}_{\rm min}$ [yN]} & \textbf{$S^{\rm q}$ [yN$\cdot$Hz$^{-1/2}$]} & \textbf{$F^{\rm q+c}_{\rm min}$ [yN]} & \textbf{$S^{\rm q+c}$ [yN$\cdot$Hz$^{-1/2}$]}\\ \hline
        2             & 0.1                                     &$1\cdot 10^{-3}$     & 9.66  & 0.30 &80.9   & 2.48    \\ \hline  2             & 1                                     &$1\cdot 10^{-4}$     & 96.6  & 0.93 &809   & 7.85    \\ \hline 2             & 10                                     &$1\cdot 10^{-5}$     & 966 & 2.96 &8090   & 24.8  \\ \hline 10             & 0.1                                     &$3\cdot 10^{-2}$   & 5.8  & 1.05 &28.5  & 5.18    \\ \hline 10            & 1                                     &$3\cdot 10^{-3}$    & 58  & 3.34 &285   & 16.4    \\ \hline 10            & 10                                   &$3\cdot 10^{-4}$  & 580  & 10.5 &2850   & 51.8   \\ \hline 30            & 0.1                                   &$1.4\cdot 10^{-1}$   & 2.23  & 0.84 &2.60  & 0.98  \\ \hline 30            & 1                                   &$1.4\cdot 10^{-2}$   & 22.3 &2.67  &26.0  & 3.11 \\ \hline 30            & 10                                   &$1.4\cdot 10^{-3}$   & 223  & 8.46 &260  & 9.84 \\ \hline

\end{tabular}
\caption{Quantum sensing results are presented for various system sizes $N$ and Coulomb coupling values $J_{\rm c}$, which represent the overall energy scale. The superscript ${\rm q}$ denotes the force and sensitivity results considering quantum shot noise, while ${\rm q+c}$ includes the effects of classical noise corresponding to a displacement of $0.2$ $\mu$m. The remaining parameters are $g/J_{\rm c}=\epsilon/J_{\rm c}=1$, $\gamma/J_{\rm c}=1.8 $, $\Delta \phi=\pi/4$, $\Delta/J_{\rm c}=0.5$, $\delta_\ff/J_{\rm c} =1.19 $.}
\label{tab:example_top_full_width}
\end{table*}

The error in the measurement of $\mathrm{s}_N$ has two contributions. Firstly, we have to consider the quantum shot noise,
$(\delta x_N)_{\rm q}$, which poses the ultimate precision limit.
\begin{eqnarray}
(\delta x_N)_{\rm q}^2 
& = & 
\langle x^2_N(t) \rangle - \langle x_N(t) \rangle^2 =
\nonumber \\
&=& \langle x_N^2 \rangle_{\epsilon_1 = 0}  \nonumber \\
&=& x^2_0 \langle  a_N^2 + (a_N^\dagger)^2 + 2 a^\dagger_N a_N + 1 \rangle_{\epsilon_1 = 0}.
\end{eqnarray}
In the last equation, we have used the fact that in our linear, Gaussian, system the quantum fluctuations around the average displacement can be calculated by considering the steady-state without force. 

In addition to the quantum shot noise, fluorescence measurements have a precision that is set by experimental limitations from optical diffraction, 
($\delta x_N)_{\rm c}$. We consider typical values in the literature \cite{streed2011imaging}, $(\delta x_N)_{\rm c} \approx$ $0.2$-$0.5$ $\mu$m, although advanced techniques may give access to the sub-diffraction limit \cite{drechsler2021optical,hasse2024phase}. According to the usual convention, we are within the quantum metrological regime as long as 
$(\delta x_N)_{\rm c} <  (\delta x_N)_{\rm q}$. 

The signal-to-noise ratio (SNR) of our quantum sensing protocol is defined as
\begin{equation}
    \text{SNR} (\delta_\ff) = \frac{\mathrm{s}_N(\delta_\ff)}{(\delta x_{N})_{\rm q} + (\delta x_N)_{\rm c}},
    \label{eq:SNR}
\end{equation}
with ${\rm SNR}(\delta_\ff) = 1$ setting the limit to detect a force. We therefore define the minimum force that is detectable by our sensor, $F_{\rm min}$, as the force such that SNR $= 1$. 
Since $s_N$ is linear in the applied force, we can express 
$F_{\rm min}$ as
\begin{eqnarray}
F_{\rm min} = \frac{(\delta x_{N})_{\rm q} + (\delta x_N)_{\rm c}}{\partial s_N / \delta F } ,
\end{eqnarray}
where the derivative can be calculated by using Eq. \eqref{sigmaxN.resonant} or Eq. \eqref{sigmaxN.nonresonant}, together with the relation between $\epsilon_1$ and $F_1$, obtained from Eq. \eqref{eq:epsilon},
\begin{equation}
\frac{\partial s_N(\delta_\ff)}{\partial F_1} =
\frac{x_0^2}{\sqrt{2}}\sqrt{|G_{N1}(\delta_\ff)|^2+|\bar{G}_{N1}(-\delta_\ff)|^2}.
\end{equation}
and the equivalent expression for $\delta_\ff = 0$.

\begin{figure*}[h]
    \centering
    \includegraphics[width=\textwidth]{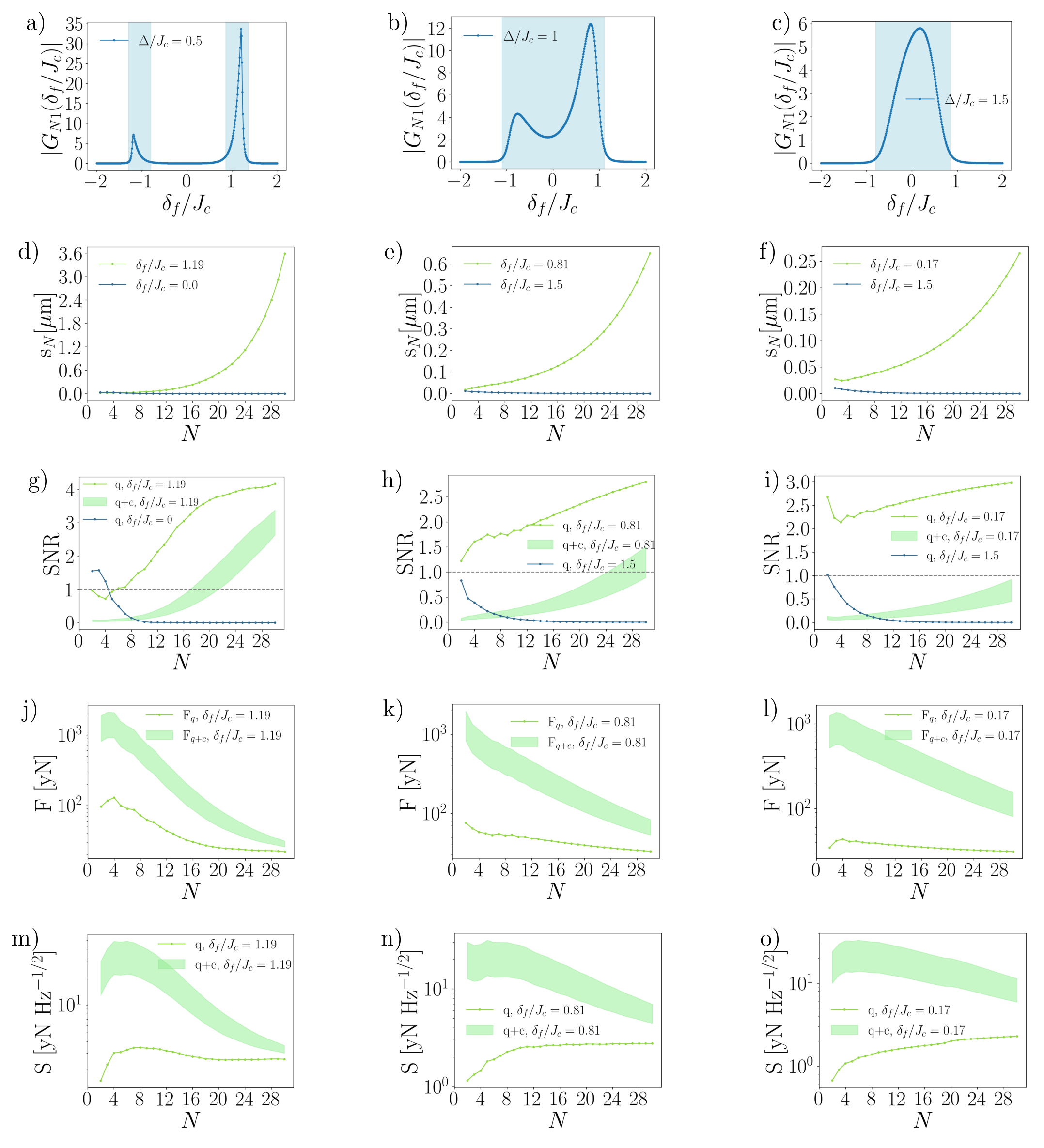}
    \caption{
    First row (a-c): Response on the $N$'th ion when the field acts on the first site as a function of the signal's detuning $\delta_\ff/J_{\rm c}$, for $\Delta/J_{\rm c}=0.5$ (a), $1$ (b), $1.5$ (c), respectively. The bluish areas correspond to topological regions.
Second row (d-f): Amplitude of oscillation in $\mu$m as a function of the number of sites for $\Delta/J_{\rm c}=0.5$ (d), $1$ (e), $1.5$ (f), respectively. We have chosen two values of $\delta_\ff/J_{\rm c}$: the topological maximum response (green) and a trivial contribution (blue). The experimental resolution limit is between $0.2-0.5$ $\mu$m.
Third row (g-i): Green dots show the SNR of the ion's displacement, considering just the quantum shot noise for $\Delta/J_{\rm c}=0.5$ (g), $1$ (h), $1.5$ (i), respectively. The shaded green area shows the SNR when considering also classical noise limited by the spatial resolution.
Fourth row (j-l): Smallest detectable force in yN as a function of the number of sites for $\Delta/J_{\rm c}=0.5$ (j), $1$ (k), $1.5$ (l), respectively. The shaded green area shows the minimal force when considering also classical noise limited by the spatial resolution.
Fifth row (m-o): Sensitivity in yN Hz$^{-1/2}$ as a function of the number of sites. The green points are related to the quantum shot noise, while the shaded green area shows the sensitivity when considering also classical noise limited by the spatial resolution. The remaining fixed parameters are $J_{\rm c}=g_s=\epsilon=1(2\pi)$ kHz, $\gamma/J_{\rm c}=1.8$, $\Delta \phi=\pi/4$, and $N=20$.}
    \label{sch23}
\end{figure*}

The impact of topology on the sensing capabilities of our system is shown in Fig. \ref{sch23}, where we present numerical results for different system sizes and values of $\Delta$ and $\delta_{\rm f}$. 
In Fig. \ref{sch23} (a-c), we plot the Green's function as a function of the force detuning $\delta_{\rm f}$, highlighting the frequency regions where topological amplification occurs. 
Fig. \ref{sch23} (d-f) shows the averaged displacement $s_N$ for values within or out of the topological region, showing the enhancement induced by topological amplification. 
Fig. \ref{sch23} (g-i) presents values of the SNR, where we observe a general monotonic growth of the SNR within the topological region. In contrast, non-topological regimes exhibit a decreasing SNR as the system size grows. 
This is a remarkable result since topological amplification enhances both the coherent signal and the noise. However, the overall balance between the two is favorable for the SNR at long lenghts ($N > 8$). 
The minimum detectable force, $F_{\rm min}$ is plotted in Fig. \ref{sch23} (j-l), where we observe, again, a monotonic enhancement in precision as $N$ increases.

\begin{figure}[h]
    \centering
    \includegraphics[width=0.5\textwidth]{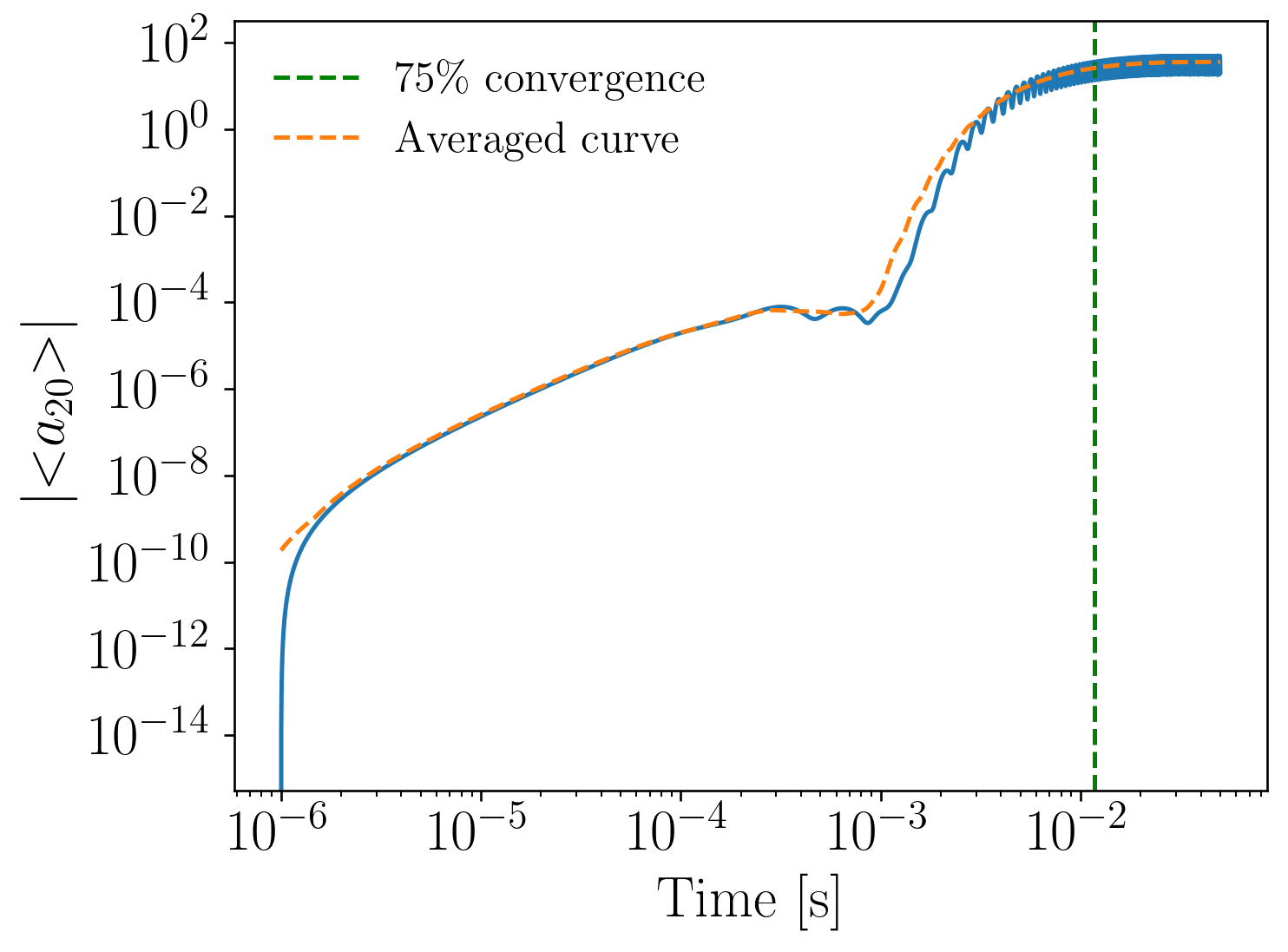}
    \caption{Time dependence of the coherence at the 20'th ion site in logarithmic scale. The steady-state oscillates around a fixed value, which we obtained by averaging over the oscillations. We fixed the criterion for steady-state formation when the time is such that the curve is within $75\%$ of the steady-state average displacement. The remaining parameters are $J_{\rm c}= 1(2\pi) $ kHz, $g/J_{\rm c} = 1$ $\Delta/J_{\rm c}=0.5$, $\delta_\ff/J_{\rm c}=1.19 $, $\Delta \phi=\pi/4$.}
    \label{timesensing}
\end{figure}

Finally, to rigorously assess the potential of our system for precision measurements, we need to account for the measurement duration. This is typically quantified by the sensitivity, $S$, defined in terms of the precision in the force measurement, $F_{\rm min}$, and the measurement time, $\tau$. 
\begin{equation}
S = F_{\rm min} \sqrt{\tau}.
\end{equation}
The rationale behind this definition is the following \cite{degen2017quantum}. Consider a number of experimental repetitions, $N_{\rm m}$, and assume that the preparation 
and readout time is negligible compared to the system's response time to the external force. 
Then, the final precision of the force measurement is  $\delta F = F_{\rm min}/\sqrt{N_{\rm m}}$.
Since $N_{\rm m} = T/\tau$, with $\tau$ being the experimental time, we find that $\delta F = S /\sqrt{T}$, with $S = F_{\rm min} \sqrt{\tau}$. Up to this point, our focus has primarily been on the steady-state behavior of the system. However, by solving the master equation, we also gain access to the time-dependent solution of the model. An example of such a calculation is shown in Fig. \ref{timesensing}. To set a criterion for convergence to the steady-state, we define the experimental time $\tau$ as the time required to reach $75\%$ of the steady-state displacement.

Fig. \ref{sch23}(m-o) shows the sensitivity scaling with $N$. 
In the full metrological regime, in which the shot noise sets the ion's position measurement error, we find that the sensitivity is better for small ion chains. 
As the system size increases, the sensitivity reaches a large $N$ limit, 
Fig. \ref{sch23}(m-n), or weakly increases with $N$, Fig. \ref{sch23}(o). 
Since $F_{\rm min}$ in Fig. \ref{sch23}(m-o) decreases with $N$, the sensitivity scaling is explained by longer relaxation times for larger $N$ leading to larger (worse) sensitivities. 
Thus, long ion chains have no advantage over $N = 2$ ion detectors in the quantum metrological regime. 
However, achieving the quantum metrological regime is also increasingly difficult for small ion chains. For example, in Fig. \ref{sch23}(m-o), we observe that average displacements $s_N$ for $N = 2$ are below $0.05$  $\mu$m. 
Thus, topological amplification is advantageous in a non-quantum metrological limit, in which experimental error limits the precision of the displacement measurement.
We address that limit in Fig. \ref{sch23}(m-o), where we plot the sensitivity for $(\delta x)_{\rm c}$ in the range $0.2-0.5$ $\mu$m, and find a monotonic enhancement of the sensitivity with $N$.

We summarize our results in Table I, with the general frequency scale set by values of $J_{\rm c} = $0.1, 1, and 10 kHz. The best sensitivity result is obtained in the quantum metrological regime, $S^{\rm q} =$ 0.3 yN $\cdot$ Hz$^{-1/2}$, with $N = 2$ ions and weak couplings $J_{\rm c} = 0.1$ kHz.
If we consider a non-quantum metrological limit with measurement error $(\delta x)_{\rm c} = 0.2$ $\mu$m, the best performance is for a chain of $N = 30$ ions and  
$J_{\rm c} = $0.1 kHz, leading to $S^{\rm q + c} =$ 0.98 yN $\cdot$ Hz$^{-1/2}$.

A wide range of experimental setups can achieve ultra-weak force sensitivities, ranging from neutral atoms \cite{degen2017quantum} to levitated nanoparticles \cite{liang2023yoctonewton}. Our results show a sensing performance in the yN $\cdot$ Hz$^{-1/2}$ regime, comparable to previous theoretical proposals involving single ions \cite{maiwald2009stylus,ivanov2016high}. A remarkable aspect of our proposal is that the frequency spectral range over which forces can be detected (the shaded topological regions in Fig. \ref{sch23} (a-c)) can be tuned by adjusting $\Delta$, which, in turn, can be controlled by adjusting the frequency of the periodic drive, $\omega_{\rm d}$, without the need of altering the trapping frequency.  

\begin{figure}[h]
    \centering
    \includegraphics[width=0.5\textwidth]{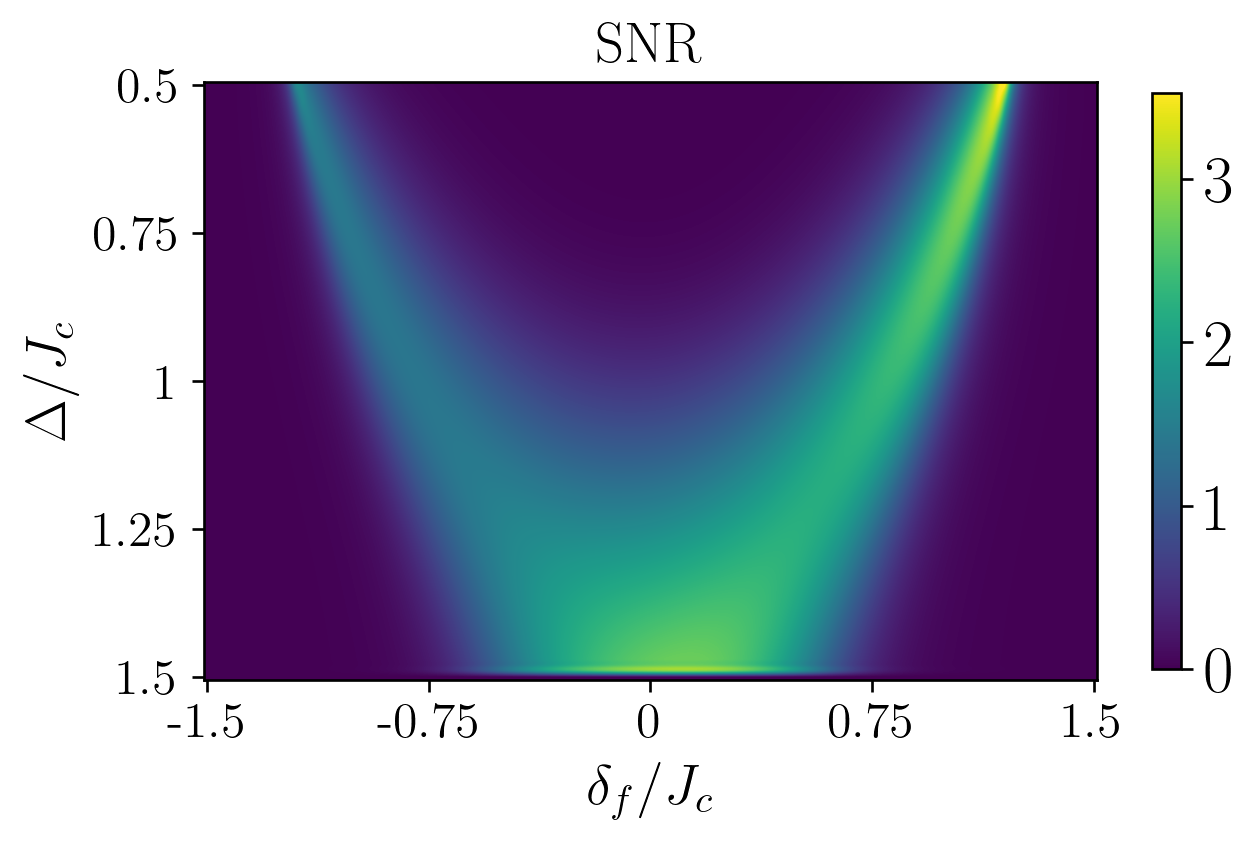}
    \caption{Parameter space spanned by detuning $\Delta/J_c$ and signal's frequency detuning $\delta_\ff/J_c$, showing the signal-to-noise ratio for $N=20$. The yellowish regions are topological with $J_{\rm c}= 1(2\pi) $ kHz, $g/J_{\rm c}=1$, $\gamma/J_{\rm c}=1.8$ and $\Delta \phi=\pi/4$. }
    \label{schemeardi}
\end{figure}

\section{Conclusions and Outlook}
\label{sec.conclusions}
We have presented a proposal for implementing topological non-equilibrium phases in a driven-dissipative trapped ion chain. The interplay between parametric driving and continuous laser cooling gives rise to non-Hermitian topological driven-dissipative phases in which the system behaves as a stable directional amplifier of vibrational excitations.  
Our proposal would lead to the realization of a topological dissipative phase in the quantum regime in a system that can be scalable to tens of sites. 

Within the same framework, we have also investigated the performance of a trapped ion chain as a quantum sensor for detecting ultra-weak forces. Our analysis predicts that the signal-to-noise ratio increases with the system size, a promising result compared to the exponential decay typically observed in trivial configurations. By measuring the displacement amplitude via fluorescence at sub-micrometer scales, we achieve force sensitivities on the order of $1$ yN $\cdot$ Hz$^{-1/2}$. This approach yields sensitivities comparable to the best results obtained in previous theoretical proposals \cite{maiwald2009stylus} while offering improved precision over prior experimental implementations \cite{biercuk2010ultrasensitive}. Moreover, the topological amplification of the signal eliminates the need for sub-diffraction limit techniques, and our proposal enhances the tunability of the frequency range for detecting external forces by solely adjusting the frequency of the periodic drive instead of modifying the trapping frequency.

As an outlook, extending the model to higher dimensions would be promising, for example, by using topological edge-states as chiral channels for amplification \cite{peano2016topological,vega2024arXiv} or defining topological invariants for dimensions higher than one. 
Further improvements of the sensitivity can achieved by, for example, optimizing the evolution time. Our proposal also leads to an exciting platform for the quantum simulation of non-equilibrium interacting topological systems by including non-linearities such as spin-phonon couplings.

\section*{Acknowledgements}
We thank Tobias Schaetz and Ulrich Warring for many fruitful discussions. 
We acknowledge support from Spanish projects PID2021-127968NBI00 funded by MICIU/AEI/10.13039/501100011033,
by Proyecto Sin\'ergico CAM 2020 Y2020/TCS-6545
(NanoQuCo-CM), and the CSIC Research Platform on Quantum Technologies PTI-001. T. Ramos
further acknowledges support from the Ramón y
Cajal program RYC2021-032473-I, financed by
MCIN/AEI/10.13039/501100011033 and the European Union NextGenerationEU/PRTR.

\begin{appendices}

\section{Continuous laser cooling}\label{app:anexoA}
In this section we derive the continuous laser cooling Liouvillian \eqref{dissipation} 
by using the theoretical framework first presented in Ref. \cite{cirac1992laser}. 
The fundamental idea is the adiabatic elimination of a fast decaying qubit by applying perturbation theory in Liouville space.
We discuss the derivation with a single ion, although our derivation can be extended to many ions, provided the conditions for the adiabatic elimination are satisfied.

We consider two internal electronic levels in a single trapped ion, 
$| 0 \rangle$, 
$| 1 \rangle$, which will act as a reservoir. 
We assume fast radiative decay from $|1\rangle$ into  $|0\rangle$ with rate $\gamma_{\rm d}$. 
The internal levels are coupled to one of the ion's vibrational modes by a red-detuned laser. The single ion is described by the master equation
\begin{equation}
    \frac{d{\rho}}{dt} =\mathcal{L}_{\rm I}(\rho) + \mathcal{L}_{\gamma_{\rm d}}({\rho}),
\end{equation}
which contains the spin-phonon interaction induced by the laser,
$\mathcal{L}_{\rm I}(\rho) = - i \left[ {H_{\rm I}}, \rho \right]$,  
with 
\begin{equation}
H_{\rm I} = g_{\rm r}( \sigma_+ a + \sigma_- a^\dagger) ,
\nonumber
\end{equation}
and the radiative decay term
\begin{equation}
\mathcal{L}_{\gamma_{\rm d}} (\rho) 
= \frac{\gamma_{\rm d}}{2}(2 \sigma_- \rho \sigma_+ - \sigma_+ \sigma_- \rho - \rho \sigma_+ \sigma_-)
\nonumber
\end{equation}. 

We consider the limit $\gamma_{\rm d} \gg \Omega$, such that $\mathcal{L}_{\rm I}$ can be considered a small perturbation to the fast spin decay. 

Our aim is to adiabatically eliminate the internal states and derive effective cooling dynamics. For this we define the projection superoperator in Liouville space
\begin{equation}
    \mathcal{P}\rho=\rho_a\otimes \ket{0}\bra{0},
\end{equation}
which projects into a product state of the reduced density matrix of the bosonic modes times the steady state of the qubit. Defining also the complementary projection operator $\mathcal{Q}=\mathds{1}-\mathcal{P}$, we look for a perturbative equation for the dynamics of $\mathcal{P}\rho$,
\begin{eqnarray}
    \mathcal{P}\dot{\rho} &= &\mathcal{P}\mathcal{L}_I\mathcal{Q}\rho, \\
    \mathcal{Q}\dot{\rho} &= &\mathcal{Q}\mathcal{L}_{\gamma_{\rm d}}\mathcal{Q}\rho+\mathcal{Q}\mathcal{L}_I\mathcal{P}\rho+ \mathcal{Q}\mathcal{L}_I\mathcal{Q}\rho,
\end{eqnarray}
where we have used that $\mathcal{L}_{\gamma_{\rm d}}\mathcal{P}\rho=0$ and $\mathcal{P}\mathcal{L}_I\mathcal{P}=0$. Taking use of the Nakajima-Zwanzig equation up to second order in the perturbative term, we can write,
\begin{equation}
    \mathcal{P}\dot{\rho}=\int^t_0 d\tau \mathcal{P}\mathcal{L}_I\mathcal{Q}e^{\mathcal{L}_{\gamma_{\rm d}}\tau}\mathcal{Q}\mathcal{L}_I\mathcal{P}\rho.
\end{equation}
Since we are interested in the dynamics of the bosonic modes $\dot{\rho}_a$, we should do the partial trace over the qubit degrees of freedom,
\begin{equation}\label{rdm}
    \dot{\rho}_a = \Tr_q{\int^t_0 d\tau \mathcal{P}\mathcal{L}_I\mathcal{Q}e^{\mathcal{L}_{\gamma_{\rm d}}\tau}\mathcal{Q}\mathcal{L}_I\mathcal{P}\rho}.
\end{equation}
We apply the definition of $\mathcal{P}\rho$ and $\mathcal{L}_I$ in terms of the commutators and develop the expression inside the integral, giving rise to
\begin{equation}\label{rhs2}
    \dot{\rho}_a = -\Tr_q \int^t_0 d\tau  
    \left[ H_{\rm I}, 
    \mathcal{Q}e^{\mathcal{L}_{\gamma_{\rm d}}\tau}\mathcal{Q}
    \left[
    H_{\rm I},\rho_a\otimes \ket{0}\bra{0} \right] \right] .
\end{equation}
To evaluate this expression, we need to use that
$\mathcal{Q}\sigma^{\pm}=\sigma^{\pm}$ and $\mathcal{L}_{\gamma_{\rm d}}(\sigma^\pm) = -
\frac{{\gamma_{\rm d}}}{2}\sigma^\pm$, and calculate all the terms that result from the double commutator in Eq. \eqref{rhs2}. Furthermore, we take the limit $t \to \infty$, which is well justified in the case of fast decay and we arrive at
\begin{equation}
    \dot{\rho}_a=\frac{\gamma}{2}(2a\rho_a a^\dagger -a^\dagger a \rho_a - \rho_a a^\dagger a),
\end{equation}
with the continuous cooling decay rate given by 
\begin{equation}
\gamma=\frac{4 g_{\rm r}^2}{{\gamma_{\rm d}}}.
\end{equation}
This result can be extended to many ions if the decay rate $\gamma_{\rm d}$ is much larger than any other dynamics. Typically, $\gamma_{\rm d} =$ 1 $(2 \pi)$ MHz and red-sideband rates $g_{\rm r} =$ 10 - 100 $(2 \pi)$ kHz \cite{Leibfried03rmp}, leading to cooling rates
$\gamma = $ 0.4 - 40 $(2 \pi)$ kHz, in line with the values required for nontrivial topological phases.

\section{Phonon correlation matrix and stability}
\label{app:anexoB}
To characterize stability and phonon correlations, we write down the equations of motion for the correlation matrix in Nambu notation defined in Eq. \eqref{pcf}. 
After some algebra, we arrive at the following equation in matrix form.
\begin{equation}\label{mat}
    \frac{dC}{dt} = i\mathbb{H}^* C- i C\mathbb{H}^T + D,
\end{equation}
with
\begin{equation}\label{dis}
    D_{\mu \nu} = 2\begin{pmatrix}0 & 0\\0&\Gamma\end{pmatrix} .
\end{equation}
Assuming now that $\mathbb{H}$ can be diagonalized, we write $\mathbb{H} = V \Lambda V^{-1}$, where $\Lambda$ is a diagonal matrix,  $\Lambda_{\mu \nu} = \delta_{\mu \nu} \lambda_\mu$. 
In order to solve Eq. \eqref{mat}, we define the transformed correlation matrices
$\Tilde{C}\equiv (V^{-1})^* C (V^{-1})^T$ and $\Tilde{D}\equiv (V^{-1})^* D (V^{-1})^T$, which follow the equation
\begin{equation}\label{mat}
    \frac{d\Tilde{C}_{\mu\nu}}{dt}=i\lambda_\mu^*\Tilde{C}_{\mu\nu}-i\Tilde{C}_{\mu\nu}\lambda_\nu +\Tilde{D}_{\mu \nu}.
\end{equation}
This differential equation can be readily solved,
\begin{equation}\label{solucionC}
    \Tilde{C}_{\mu \nu}(t)=\Tilde{D}_{\mu \nu}\frac{e^{i(\lambda^*_\mu-\lambda_\nu)t}-1}{i(\lambda^*_\mu-\lambda_\nu)} + C_{\mu \nu}(0)e^{i(\lambda^*_\mu-\lambda_\nu)t}.
\end{equation}
From this expression, it is clear that the condition $\Im{\lambda_\mu} < 0$ is necessary for stability. If the system is stable, then we get the steady state
\begin{equation}
    \Tilde{C}_{\mu \nu}=\frac{-\Tilde{D}_{\mu \nu}}{i(\lambda^*_\mu-\lambda_\nu)}.
    \label{Ctransformed}
\end{equation}

Eq. \eqref{Ctransformed} is equivalent to the convolution form in Eq. \eqref{eq:Cintegral}. 
To show this equivalence we write the steady-state value of the transformed matrix $\Tilde{C}_{\mu \nu}$ predicted by Eq. \eqref{eq:Cintegral},
\begin{equation}\label{112}
    \Tilde{C}_{\mu \nu}=\int \frac{d\omega}{2 \pi} \frac{1}{\omega-\lambda^*_\mu}\frac{\Tilde{D}_{\mu \nu}}{2}\frac{1}{\omega-\lambda_\nu}.
\end{equation}
The integral in Eq \eqref{112} can be solved in the complex plane. There are two simple poles at $\omega=\lambda^*_\mu$ and $\omega=\lambda_\nu$, respectively. By applying the residue theorem we get to Eq. \eqref{eq:Cintegral}. 

\section{Topological amplification theory}\label{app:anexoE}
The topological properties of our system determine the steady-state and linear response, through an effect known as topological amplification. We review here the basic mathematical formulation of this phenomenon.

The main result that we need to prove is the connection between the existence of edge singular vectors and the winding number $\nu(\omega)$ in Eq. \eqref{winding1}. 
Essential to this connection is the \textit{doubled dynamical matrix} $\mathcal{H}$ defined as 
\begin{eqnarray}
   \mathcal{H}(\omega) &\equiv &\begin{pmatrix}0 & \omega-\mathbb{H} \\\omega-\mathbb{H}^\dagger&0\end{pmatrix} .
\end{eqnarray}
${\cal H}(\omega)$ is, by construction, an Hermitian matrix, and its topological properties can be analyzed within standard topological insulator theory.
The key point of $\mathcal{H}(\omega)$ is that its eigensystem is equivalent to the singular value decomposition of $\mathbb{H}$. As can be easily checked
\begin{equation}  
\mathcal{H}\begin{pmatrix}U\\ V\end{pmatrix} 
= 
\begin{pmatrix} U S \\ V S \end{pmatrix}.
\end{equation}
This equivalence implies that the topological edge states of ${\cal H}(\omega)$ will immediately lead to zero-singular values and edge-singular vectors and, thus, directional amplification in the system. 
Another key observation is that ${\cal H}(\omega)$ has a chiral symmetry,
\begin{equation}\label{chiral}
    \mathcal{S} \mathcal{H}(\omega) \mathcal{S}^{-1} = - \cal{H}(\omega) ,
\end{equation}
with 
\begin{eqnarray}
   \mathcal{S} =
   \begin{pmatrix} \mathbb{1} & 0 \\ 0 & - \mathbb{1} \end{pmatrix} .
   \label{chiral.S}
\end{eqnarray}
This chiral symmetry protects topological edge states. 
However, we note that $\mathcal{S}$ is not a physical symmetry; instead, it holds by construction. 
Thus, the topological protection of the singular edge states is an inherent property of our system that does not depend on any physical symmetry.

In one dimension, the nontrivial symmetry classes that lead to different topological phases are characterized by a topological invariant. In particular, we will use the winding number $\nu$ as a topological invariant, which is defined in terms of 
${\cal H}(\omega)$ as \cite{gomez2023driven} 
\begin{equation}\label{winding2}
    \nu(\omega)=\int_{-\pi}^{\pi}\frac{dk}{4\pi i}\Tr\left[\tau_z \mathcal{H}^{-1}(k,\omega)\partial_k \mathcal{H}(k,\omega) \right],
\end{equation} 
where $\mathcal{H}(k,\omega)$ is the Fourier transform of the doubled matrix and $\tau_z$ is the Pauli matrix related 
to its intrinsic chiral symmetry. 
By substituting the form of the doubled dynamical matrix, we can show that the expression for $\nu(\omega)$ in Eq. \eqref{winding2} is equivalent to Eq. \eqref{winding1}.

\section{classification of topological phases}
\label{app:anex_classification}
This section reviews the appearance of topological phases in our model. Although this may seem like a very theoretical issue, it can have practical consequences in determining ranges of parameters for which topological amplification and enhanced force sensing can be achieved.

We start by re-writing our chiral doubled Hamiltonian in terms of an additional pseudo-spin degree of freedom spanned by a basis of Pauli matrices (for ease of notation, we suppress the variable $\omega$ in ${\cal H}$)
$\{\mathds{1}, \tau_x, \tau_y, \tau_z \}$,
\begin{eqnarray}\label{def}
   \mathcal{H}(k) =  (\omega-\mathbb{H}(k))\otimes \tau^+ + (\omega-\mathbb{H}(k)^\dagger) \otimes \tau^- ,
\end{eqnarray}
with the non-Hermitian dynamical matrix in the plane-wave basis,
\begin{equation}
    \mathbb{H}(k)=f_0(k)\mathds{1} + \Vec{f}(k) \cdot \Vec{\sigma}.
\end{equation}
$\sigma_j$ are the Pauli matrices associated to the Nambu index. 
We do not give an explicit form for $\vec{f}(k)$, which can be read from 
Eq. (\ref{hnh}), but rather express ${\cal H}(k)$ in terms of the basis spanned by the tensor products 
$\{\sigma_j \otimes \tau_l \}$,
\begin{align}
\mathcal{H}(k)  = &  \\
h_{0x} (\mathds{1} \! & \otimes \! \tau_x) & + h_{0y}(\mathds{1}\! \otimes \! \tau_y) 
+ h_{yy}(\sigma_y \! \otimes \! \tau_y )+ h_{zx}(\sigma_z \! \otimes \! \tau_x), \nonumber
\end{align}
with components
\begin{eqnarray}
    h_{0x} &= &\omega+2J_{\rm c}\sum_n\frac{\sin{(nk)}}{n^3}\sin{(n\Delta \phi)}, \nonumber \\
    h_{0y} &= &\frac{\gamma}{2}, \nonumber \\
    h_{yy} &= & -g,\nonumber \\
    h_{zx} &= &\Delta + 2J_{\rm c}\sum_n\frac{\cos{(nk)}}{n^3}\cos{(n\Delta \phi)}. 
    \label{eq:hij}
\end{eqnarray}

The symmetries of ${\cal H}(k)$ determine the symmetry class into which it can 
be classified according to the ten-fold way for topological insulators \cite{altland1997nonstandard}. In particular, we have to consider discrete symmetry operators for time-reversal, ${\cal T}$, and charge conjugation, ${\cal C}$, 
which can be written as
\begin{align}
{\cal T} & = U_T K , \nonumber \\
{\cal C} & = U_C K ,
\end{align}
where $U_T$, $U_C$ are unitary matrices. $K$ is the complex conjugation operator, fulfilling $K^2 = 1$ and $K i K = -i$. 
Time reversal and charge conjugation operators must fulfill the conditions ${\cal T}^2 = \pm 1$ and 
${\cal C}^2 = \pm 1$, which leads to the constraints
\begin{equation}
U_T U_T^* = \pm \mathbb{1}, \ \ U_C U_C^* = \pm \mathbb{1} .
\label{restriction}
\end{equation} 
Finally, $\mathcal{T}$ and $\mathcal{C}$ are related to the chiral symmetry $\mathcal{S}$ by 
$\mathcal{T} \mathcal{C} = \mathcal{S}$. In our system, $S$ is defined by Eq.~\eqref{chiral.S}, so that the relation
\begin{equation}
U_T U_C^* \propto \sigma_z     
\end{equation}
has to be imposed upon $U_T$, $U_C$. 

In a translationally invariant system, time-reversal and charge conjugation symmetries are fulfilled if there exist unitary matrices $U_T$ and $U_C$, such that 
$T {\cal H}(k) T^{-1} = {\cal H}(-k)$ 
and $C {\cal H}(k) C^{-1} = -{\cal H}(-k)$, leading to conditions
\begin{align}
U_T {\cal H}(k)^* U_T^\dagger = {\cal H}(-k), \nonumber \\
U_C {\cal H}(k)^* U_C^\dagger = -{\cal H}(-k) .
\end{align}
Using this formalism, we identify the following symmetry classes of possible trapped-ion parametric chains:

\begin{itemize}

\item   $\Delta \phi \neq 0,\pi$ and $\omega\neq 0$ $\to$  Class AIII  

If both $\omega$ and $\Delta \phi$ are not zero, then $h_{0x}$ in Eqs. \eqref{eq:hij} changes in magnitude under the action of the transformation $k \to -k$, leading to the breaking of time-reversal and charge conjugation symmetry, corresponding to the AIII symmetry class, which can show zero-energy edge states.

\item $\Delta \phi \neq 0,\pi$ and $\omega=0$ $\to$ Class BDI 

Here, time-reversal symmetry is implemented by $U_T=\sigma_x\otimes \tau_z$ and charge conjugation symmetry by $U_C = \sigma_x\otimes \mathds{1}$, leading to $\mathcal{T}^2 = \mathcal{C}^2 = 1$, within the BDI symmetry class, which can have zero-energy edge states.

\item $\Delta \phi=0, \pi$ $\to$ Class CI  

Time-reversal symmetry is implemented by $U_T=\sigma_z\otimes \tau_x$ that gives $\mathcal{T}^2=+1$, and charge conjugation by $U_C=\sigma_z \otimes \tau_y$, leading to $\mathcal{C}^2=-1$. This leads to the symmetry class CI, where all topological phases are trivial.

\end{itemize}

Topological insulator theory, thus, predicts that $\Delta \phi \neq 0$ is required to have nontrivial topological phases of ${\cal H}(k)$ and topological amplification.

\end{appendices}

\bibliography{topbib.bib}

\begin{thebibliography}{49}%
\makeatletter
\providecommand \@ifxundefined [1]{%
 \@ifx{#1\undefined}
}%
\providecommand \@ifnum [1]{%
 \ifnum #1\expandafter \@firstoftwo
 \else \expandafter \@secondoftwo
 \fi
}%
\providecommand \@ifx [1]{%
 \ifx #1\expandafter \@firstoftwo
 \else \expandafter \@secondoftwo
 \fi
}%
\providecommand \natexlab [1]{#1}%
\providecommand \enquote  [1]{``#1''}%
\providecommand \bibnamefont  [1]{#1}%
\providecommand \bibfnamefont [1]{#1}%
\providecommand \citenamefont [1]{#1}%
\providecommand \href@noop [0]{\@secondoftwo}%
\providecommand \href [0]{\begingroup \@sanitize@url \@href}%
\providecommand \@href[1]{\@@startlink{#1}\@@href}%
\providecommand \@@href[1]{\endgroup#1\@@endlink}%
\providecommand \@sanitize@url [0]{\catcode `\\12\catcode `\$12\catcode `\&12\catcode `\#12\catcode `\^12\catcode `\_12\catcode `\%12\relax}%
\providecommand \@@startlink[1]{}%
\providecommand \@@endlink[0]{}%
\providecommand \url  [0]{\begingroup\@sanitize@url \@url }%
\providecommand \@url [1]{\endgroup\@href {#1}{\urlprefix }}%
\providecommand \urlprefix  [0]{URL }%
\providecommand \Eprint [0]{\href }%
\providecommand \doibase [0]{https://doi.org/}%
\providecommand \selectlanguage [0]{\@gobble}%
\providecommand \bibinfo  [0]{\@secondoftwo}%
\providecommand \bibfield  [0]{\@secondoftwo}%
\providecommand \translation [1]{[#1]}%
\providecommand \BibitemOpen [0]{}%
\providecommand \bibitemStop [0]{}%
\providecommand \bibitemNoStop [0]{.\EOS\space}%
\providecommand \EOS [0]{\spacefactor3000\relax}%
\providecommand \BibitemShut  [1]{\csname bibitem#1\endcsname}%
\let\auto@bib@innerbib\@empty
\bibitem [{\citenamefont {Leibfried}\ \emph {et~al.}(2003)\citenamefont {Leibfried}, \citenamefont {Blatt}, \citenamefont {Monroe},\ and\ \citenamefont {Wineland}}]{Leibfried03rmp}%
  \BibitemOpen
  \bibfield  {author} {\bibinfo {author} {\bibfnamefont {D.}~\bibnamefont {Leibfried}}, \bibinfo {author} {\bibfnamefont {R.}~\bibnamefont {Blatt}}, \bibinfo {author} {\bibfnamefont {C.}~\bibnamefont {Monroe}},\ and\ \bibinfo {author} {\bibfnamefont {D.}~\bibnamefont {Wineland}},\ }\bibfield  {title} {\bibinfo {title} {Quantum dynamics of single trapped ions},\ }\href {https://doi.org/10.1103/RevModPhys.75.281} {\bibfield  {journal} {\bibinfo  {journal} {Rev. Mod. Phys.}\ }\textbf {\bibinfo {volume} {75}},\ \bibinfo {pages} {281} (\bibinfo {year} {2003})}\BibitemShut {NoStop}%
\bibitem [{\citenamefont {Blatt}\ and\ \citenamefont {Roos}(2012)}]{blatt2012quantum}%
  \BibitemOpen
  \bibfield  {author} {\bibinfo {author} {\bibfnamefont {R.}~\bibnamefont {Blatt}}\ and\ \bibinfo {author} {\bibfnamefont {C.~F.}\ \bibnamefont {Roos}},\ }\bibfield  {title} {\bibinfo {title} {Quantum simulations with trapped ions},\ }\href {https://www.nature.com/articles/nphys2252} {\bibfield  {journal} {\bibinfo  {journal} {Nat. Phys.}\ }\textbf {\bibinfo {volume} {8}},\ \bibinfo {pages} {277} (\bibinfo {year} {2012})}\BibitemShut {NoStop}%
\bibitem [{\citenamefont {Schneider}\ \emph {et~al.}(2012)\citenamefont {Schneider}, \citenamefont {Porras},\ and\ \citenamefont {Schaetz}}]{schneider2012experimental}%
  \BibitemOpen
  \bibfield  {author} {\bibinfo {author} {\bibfnamefont {C.}~\bibnamefont {Schneider}}, \bibinfo {author} {\bibfnamefont {D.}~\bibnamefont {Porras}},\ and\ \bibinfo {author} {\bibfnamefont {T.}~\bibnamefont {Schaetz}},\ }\bibfield  {title} {\bibinfo {title} {Experimental quantum simulations of many-body physics with trapped ions},\ }\href {https://iopscience.iop.org/article/10.1088/0034-4885/75/2/024401/meta?casa_token=DuR1PtcqBVoAAAAA:3pUVgSq0cZ4L1cGTARLUU_vsrAndwXC6St1VUPiwNVvD_CrCZgGj1OU0SaI32xRqwJe0zTzny-eZHhsk8FfXTnTvfj2o} {\bibfield  {journal} {\bibinfo  {journal} {Rep. Prog. Phys.}\ }\textbf {\bibinfo {volume} {75}},\ \bibinfo {pages} {024401} (\bibinfo {year} {2012})}\BibitemShut {NoStop}%
\bibitem [{\citenamefont {Monroe}\ \emph {et~al.}(2021)\citenamefont {Monroe}, \citenamefont {Campbell}, \citenamefont {Duan}, \citenamefont {Gong}, \citenamefont {Gorshkov}, \citenamefont {Hess}, \citenamefont {Islam}, \citenamefont {Kim}, \citenamefont {Linke}, \citenamefont {Pagano}, \citenamefont {Richerme}, \citenamefont {Senko},\ and\ \citenamefont {Yao}}]{monroe21review}%
  \BibitemOpen
  \bibfield  {author} {\bibinfo {author} {\bibfnamefont {C.}~\bibnamefont {Monroe}}, \bibinfo {author} {\bibfnamefont {W.~C.}\ \bibnamefont {Campbell}}, \bibinfo {author} {\bibfnamefont {L.-M.}\ \bibnamefont {Duan}}, \bibinfo {author} {\bibfnamefont {Z.-X.}\ \bibnamefont {Gong}}, \bibinfo {author} {\bibfnamefont {A.~V.}\ \bibnamefont {Gorshkov}}, \bibinfo {author} {\bibfnamefont {P.~W.}\ \bibnamefont {Hess}}, \bibinfo {author} {\bibfnamefont {R.}~\bibnamefont {Islam}}, \bibinfo {author} {\bibfnamefont {K.}~\bibnamefont {Kim}}, \bibinfo {author} {\bibfnamefont {N.~M.}\ \bibnamefont {Linke}}, \bibinfo {author} {\bibfnamefont {G.}~\bibnamefont {Pagano}}, \bibinfo {author} {\bibfnamefont {P.}~\bibnamefont {Richerme}}, \bibinfo {author} {\bibfnamefont {C.}~\bibnamefont {Senko}},\ and\ \bibinfo {author} {\bibfnamefont {N.~Y.}\ \bibnamefont {Yao}},\ }\bibfield  {title} {\bibinfo {title} {Programmable quantum simulations of spin systems with trapped ions},\ }\href {https://doi.org/10.1103/RevModPhys.93.025001}
  {\bibfield  {journal} {\bibinfo  {journal} {Rev. Mod. Phys.}\ }\textbf {\bibinfo {volume} {93}},\ \bibinfo {pages} {025001} (\bibinfo {year} {2021})}\BibitemShut {NoStop}%
\bibitem [{\citenamefont {Cirac}\ \emph {et~al.}(1992)\citenamefont {Cirac}, \citenamefont {Blatt}, \citenamefont {Zoller},\ and\ \citenamefont {Phillips}}]{cirac1992laser}%
  \BibitemOpen
  \bibfield  {author} {\bibinfo {author} {\bibfnamefont {J.~I.}\ \bibnamefont {Cirac}}, \bibinfo {author} {\bibfnamefont {R.}~\bibnamefont {Blatt}}, \bibinfo {author} {\bibfnamefont {P.}~\bibnamefont {Zoller}},\ and\ \bibinfo {author} {\bibfnamefont {W.~D.}\ \bibnamefont {Phillips}},\ }\bibfield  {title} {\bibinfo {title} {Laser cooling of trapped ions in a standing wave},\ }\href {https://journals.aps.org/pra/abstract/10.1103/PhysRevA.46.2668} {\bibfield  {journal} {\bibinfo  {journal} {Phys. Rev. A}\ }\textbf {\bibinfo {volume} {46}},\ \bibinfo {pages} {2668} (\bibinfo {year} {1992})}\BibitemShut {NoStop}%
\bibitem [{\citenamefont {Biercuk}\ \emph {et~al.}(2010)\citenamefont {Biercuk}, \citenamefont {Uys}, \citenamefont {Britton}, \citenamefont {VanDevender},\ and\ \citenamefont {Bollinger}}]{biercuk2010ultrasensitive}%
  \BibitemOpen
  \bibfield  {author} {\bibinfo {author} {\bibfnamefont {M.~J.}\ \bibnamefont {Biercuk}}, \bibinfo {author} {\bibfnamefont {H.}~\bibnamefont {Uys}}, \bibinfo {author} {\bibfnamefont {J.~W.}\ \bibnamefont {Britton}}, \bibinfo {author} {\bibfnamefont {A.~P.}\ \bibnamefont {VanDevender}},\ and\ \bibinfo {author} {\bibfnamefont {J.~J.}\ \bibnamefont {Bollinger}},\ }\bibfield  {title} {\bibinfo {title} {Ultrasensitive detection of force and displacement using trapped ions},\ }\href {https://www.nature.com/articles/nnano.2010.165} {\bibfield  {journal} {\bibinfo  {journal} {Nat. Nanotechnol.}\ }\textbf {\bibinfo {volume} {5}},\ \bibinfo {pages} {646} (\bibinfo {year} {2010})}\BibitemShut {NoStop}%
\bibitem [{\citenamefont {Ivanov}(2020)}]{ivanov2020steady}%
  \BibitemOpen
  \bibfield  {author} {\bibinfo {author} {\bibfnamefont {P.~A.}\ \bibnamefont {Ivanov}},\ }\bibfield  {title} {\bibinfo {title} {Steady-state force sensing with single trapped ion},\ }\href {https://iopscience.iop.org/article/10.1088/1402-4896/ab444c} {\bibfield  {journal} {\bibinfo  {journal} {Phys. Scr.}\ }\textbf {\bibinfo {volume} {95}},\ \bibinfo {pages} {025103} (\bibinfo {year} {2020})}\BibitemShut {NoStop}%
\bibitem [{\citenamefont {Gilmore}\ \emph {et~al.}(2021)\citenamefont {Gilmore}, \citenamefont {Affolter}, \citenamefont {Lewis-Swan}, \citenamefont {J.}, \citenamefont {Barberena}, \citenamefont {Jordan}, \citenamefont {Rey},\ and\ \citenamefont {Bollinger}}]{bollinger_dm}%
  \BibitemOpen
  \bibfield  {author} {\bibinfo {author} {\bibfnamefont {K.~A.}\ \bibnamefont {Gilmore}}, \bibinfo {author} {\bibfnamefont {M.}~\bibnamefont {Affolter}}, \bibinfo {author} {\bibnamefont {Lewis-Swan}}, \bibinfo {author} {\bibfnamefont {R.}~\bibnamefont {J.}}, \bibinfo {author} {\bibfnamefont {D.}~\bibnamefont {Barberena}}, \bibinfo {author} {\bibfnamefont {E.}~\bibnamefont {Jordan}}, \bibinfo {author} {\bibfnamefont {A.~M.}\ \bibnamefont {Rey}},\ and\ \bibinfo {author} {\bibfnamefont {J.~J.}\ \bibnamefont {Bollinger}},\ }\bibfield  {title} {\bibinfo {title} {Quantum-enhanced sensing of displacements and electric fields with large trapped-ion crystals},\ }\href {https://doi.org/10.1126/science.abi5226} {\bibfield  {journal} {\bibinfo  {journal} {Science}\ }\textbf {\bibinfo {volume} {373}},\ \bibinfo {pages} {673} (\bibinfo {year} {2021})}\BibitemShut {NoStop}%
\bibitem [{\citenamefont {Porras}\ and\ \citenamefont {Cirac}(2004)}]{porras2004bose}%
  \BibitemOpen
  \bibfield  {author} {\bibinfo {author} {\bibfnamefont {D.}~\bibnamefont {Porras}}\ and\ \bibinfo {author} {\bibfnamefont {J.~I.}\ \bibnamefont {Cirac}},\ }\bibfield  {title} {\bibinfo {title} {Bose-einstein condensation and strong-correlation behavior of phonons in ion traps},\ }\href {https://journals.aps.org/prl/abstract/10.1103/PhysRevLett.93.263602} {\bibfield  {journal} {\bibinfo  {journal} {Phys. Rev. Lett.}\ }\textbf {\bibinfo {volume} {93}},\ \bibinfo {pages} {263602} (\bibinfo {year} {2004})}\BibitemShut {NoStop}%
\bibitem [{\citenamefont {Deng}\ \emph {et~al.}(2008)\citenamefont {Deng}, \citenamefont {Porras},\ and\ \citenamefont {Cirac}}]{Deng2008pra}%
  \BibitemOpen
  \bibfield  {author} {\bibinfo {author} {\bibfnamefont {X.-L.}\ \bibnamefont {Deng}}, \bibinfo {author} {\bibfnamefont {D.}~\bibnamefont {Porras}},\ and\ \bibinfo {author} {\bibfnamefont {J.~I.}\ \bibnamefont {Cirac}},\ }\bibfield  {title} {\bibinfo {title} {Quantum phases of interacting phonons in ion traps},\ }\href {https://doi.org/10.1103/PhysRevA.77.033403} {\bibfield  {journal} {\bibinfo  {journal} {Phys. Rev. A}\ }\textbf {\bibinfo {volume} {77}},\ \bibinfo {pages} {033403} (\bibinfo {year} {2008})}\BibitemShut {NoStop}%
\bibitem [{\citenamefont {Haze}\ \emph {et~al.}(2012)\citenamefont {Haze}, \citenamefont {Tateishi}, \citenamefont {Noguchi}, \citenamefont {Toyoda},\ and\ \citenamefont {Urabe}}]{Haze12pra}%
  \BibitemOpen
  \bibfield  {author} {\bibinfo {author} {\bibfnamefont {S.}~\bibnamefont {Haze}}, \bibinfo {author} {\bibfnamefont {Y.}~\bibnamefont {Tateishi}}, \bibinfo {author} {\bibfnamefont {A.}~\bibnamefont {Noguchi}}, \bibinfo {author} {\bibfnamefont {K.}~\bibnamefont {Toyoda}},\ and\ \bibinfo {author} {\bibfnamefont {S.}~\bibnamefont {Urabe}},\ }\bibfield  {title} {\bibinfo {title} {Observation of phonon hopping in radial vibrational modes of trapped ions},\ }\href {https://doi.org/10.1103/PhysRevA.85.031401} {\bibfield  {journal} {\bibinfo  {journal} {Phys. Rev. A}\ }\textbf {\bibinfo {volume} {85}},\ \bibinfo {pages} {031401} (\bibinfo {year} {2012})}\BibitemShut {NoStop}%
\bibitem [{\citenamefont {Toyoda}\ \emph {et~al.}(2015)\citenamefont {Toyoda}, \citenamefont {Hiji}, \citenamefont {Noguchi},\ and\ \citenamefont {Urabe}}]{Toyoda2015nat}%
  \BibitemOpen
  \bibfield  {author} {\bibinfo {author} {\bibfnamefont {K.}~\bibnamefont {Toyoda}}, \bibinfo {author} {\bibfnamefont {R.}~\bibnamefont {Hiji}}, \bibinfo {author} {\bibfnamefont {A.}~\bibnamefont {Noguchi}},\ and\ \bibinfo {author} {\bibfnamefont {S.}~\bibnamefont {Urabe}},\ }\bibfield  {title} {\bibinfo {title} {Hong--ou--mandel interference of two phonons in trapped ions},\ }\href {https://doi.org/10.1038/nature15735} {\bibfield  {journal} {\bibinfo  {journal} {Nature}\ }\textbf {\bibinfo {volume} {527}},\ \bibinfo {pages} {74} (\bibinfo {year} {2015})}\BibitemShut {NoStop}%
\bibitem [{\citenamefont {Bermudez}\ \emph {et~al.}(2011)\citenamefont {Bermudez}, \citenamefont {Schaetz},\ and\ \citenamefont {Porras}}]{bermudez_synthetic_2011}%
  \BibitemOpen
  \bibfield  {author} {\bibinfo {author} {\bibfnamefont {A.}~\bibnamefont {Bermudez}}, \bibinfo {author} {\bibfnamefont {T.}~\bibnamefont {Schaetz}},\ and\ \bibinfo {author} {\bibfnamefont {D.}~\bibnamefont {Porras}},\ }\bibfield  {title} {\bibinfo {title} {Synthetic {Gauge} {Fields} for {Vibrational} {Excitations} of {Trapped} {Ions}},\ }\href {https://doi.org/10.1103/PhysRevLett.107.150501} {\bibfield  {journal} {\bibinfo  {journal} {Phys. Rev. Lett.}\ }\textbf {\bibinfo {volume} {107}},\ \bibinfo {pages} {150501} (\bibinfo {year} {2011})}\BibitemShut {NoStop}%
\bibitem [{\citenamefont {Kiefer}\ \emph {et~al.}(2019)\citenamefont {Kiefer}, \citenamefont {Hakelberg}, \citenamefont {Wittemer}, \citenamefont {Berm{\'u}dez}, \citenamefont {Porras}, \citenamefont {Warring},\ and\ \citenamefont {Schaetz}}]{kiefer2019floquet}%
  \BibitemOpen
  \bibfield  {author} {\bibinfo {author} {\bibfnamefont {P.}~\bibnamefont {Kiefer}}, \bibinfo {author} {\bibfnamefont {F.}~\bibnamefont {Hakelberg}}, \bibinfo {author} {\bibfnamefont {M.}~\bibnamefont {Wittemer}}, \bibinfo {author} {\bibfnamefont {A.}~\bibnamefont {Berm{\'u}dez}}, \bibinfo {author} {\bibfnamefont {D.}~\bibnamefont {Porras}}, \bibinfo {author} {\bibfnamefont {U.}~\bibnamefont {Warring}},\ and\ \bibinfo {author} {\bibfnamefont {T.}~\bibnamefont {Schaetz}},\ }\bibfield  {title} {\bibinfo {title} {Floquet-engineered vibrational dynamics in a two-dimensional array of trapped ions},\ }\href {https://journals.aps.org/prl/abstract/10.1103/PhysRevLett.123.213605} {\bibfield  {journal} {\bibinfo  {journal} {Phys. Rev. Lett.}\ }\textbf {\bibinfo {volume} {123}},\ \bibinfo {pages} {213605} (\bibinfo {year} {2019})}\BibitemShut {NoStop}%
\bibitem [{\citenamefont {Porras}\ \emph {et~al.}(2008)\citenamefont {Porras}, \citenamefont {Marquardt}, \citenamefont {von Delft},\ and\ \citenamefont {Cirac}}]{Porras2008pra}%
  \BibitemOpen
  \bibfield  {author} {\bibinfo {author} {\bibfnamefont {D.}~\bibnamefont {Porras}}, \bibinfo {author} {\bibfnamefont {F.}~\bibnamefont {Marquardt}}, \bibinfo {author} {\bibfnamefont {J.}~\bibnamefont {von Delft}},\ and\ \bibinfo {author} {\bibfnamefont {J.~I.}\ \bibnamefont {Cirac}},\ }\bibfield  {title} {\bibinfo {title} {Mesoscopic spin-boson models of trapped ions},\ }\href {https://doi.org/10.1103/PhysRevA.78.010101} {\bibfield  {journal} {\bibinfo  {journal} {Phys. Rev. A}\ }\textbf {\bibinfo {volume} {78}},\ \bibinfo {pages} {010101} (\bibinfo {year} {2008})}\BibitemShut {NoStop}%
\bibitem [{\citenamefont {Ivanov}\ \emph {et~al.}(2009)\citenamefont {Ivanov}, \citenamefont {Ivanov}, \citenamefont {Vitanov}, \citenamefont {Mering}, \citenamefont {Fleischhauer},\ and\ \citenamefont {Singer}}]{Ivanov2009pra}%
  \BibitemOpen
  \bibfield  {author} {\bibinfo {author} {\bibfnamefont {P.~A.}\ \bibnamefont {Ivanov}}, \bibinfo {author} {\bibfnamefont {S.~S.}\ \bibnamefont {Ivanov}}, \bibinfo {author} {\bibfnamefont {N.~V.}\ \bibnamefont {Vitanov}}, \bibinfo {author} {\bibfnamefont {A.}~\bibnamefont {Mering}}, \bibinfo {author} {\bibfnamefont {M.}~\bibnamefont {Fleischhauer}},\ and\ \bibinfo {author} {\bibfnamefont {K.}~\bibnamefont {Singer}},\ }\bibfield  {title} {\bibinfo {title} {Simulation of a quantum phase transition of polaritons with trapped ions},\ }\href {https://doi.org/10.1103/PhysRevA.80.060301} {\bibfield  {journal} {\bibinfo  {journal} {Phys. Rev. A}\ }\textbf {\bibinfo {volume} {80}},\ \bibinfo {pages} {060301} (\bibinfo {year} {2009})}\BibitemShut {NoStop}%
\bibitem [{\citenamefont {Debnath}\ \emph {et~al.}(2018)\citenamefont {Debnath}, \citenamefont {Linke}, \citenamefont {Wang}, \citenamefont {Figgatt}, \citenamefont {Landsman}, \citenamefont {Duan},\ and\ \citenamefont {Monroe}}]{Monroe2018prl}%
  \BibitemOpen
  \bibfield  {author} {\bibinfo {author} {\bibfnamefont {S.}~\bibnamefont {Debnath}}, \bibinfo {author} {\bibfnamefont {N.~M.}\ \bibnamefont {Linke}}, \bibinfo {author} {\bibfnamefont {S.-T.}\ \bibnamefont {Wang}}, \bibinfo {author} {\bibfnamefont {C.}~\bibnamefont {Figgatt}}, \bibinfo {author} {\bibfnamefont {K.~A.}\ \bibnamefont {Landsman}}, \bibinfo {author} {\bibfnamefont {L.-M.}\ \bibnamefont {Duan}},\ and\ \bibinfo {author} {\bibfnamefont {C.}~\bibnamefont {Monroe}},\ }\bibfield  {title} {\bibinfo {title} {Observation of hopping and blockade of bosons in a trapped ion spin chain},\ }\href {https://doi.org/10.1103/PhysRevLett.120.073001} {\bibfield  {journal} {\bibinfo  {journal} {Phys. Rev. Lett.}\ }\textbf {\bibinfo {volume} {120}},\ \bibinfo {pages} {073001} (\bibinfo {year} {2018})}\BibitemShut {NoStop}%
\bibitem [{\citenamefont {Ohira}\ \emph {et~al.}(2021)\citenamefont {Ohira}, \citenamefont {Kume}, \citenamefont {Takahashi},\ and\ \citenamefont {Toyoda}}]{OhiraQST2021}%
  \BibitemOpen
  \bibfield  {author} {\bibinfo {author} {\bibfnamefont {R.}~\bibnamefont {Ohira}}, \bibinfo {author} {\bibfnamefont {S.}~\bibnamefont {Kume}}, \bibinfo {author} {\bibfnamefont {H.}~\bibnamefont {Takahashi}},\ and\ \bibinfo {author} {\bibfnamefont {K.}~\bibnamefont {Toyoda}},\ }\bibfield  {title} {\bibinfo {title} {Polariton blockade in the jaynes–cummings–hubbard model with trapped ions},\ }\href {https://dx.doi.org/10.1088/2058-9565/abecd1} {\bibfield  {journal} {\bibinfo  {journal} {Quantum Sci. Technol.}\ }\textbf {\bibinfo {volume} {6}},\ \bibinfo {pages} {024015} (\bibinfo {year} {2021})}\BibitemShut {NoStop}%
\bibitem [{\citenamefont {Katz}\ and\ \citenamefont {Monroe}(2023)}]{Katz2023prl}%
  \BibitemOpen
  \bibfield  {author} {\bibinfo {author} {\bibfnamefont {O.}~\bibnamefont {Katz}}\ and\ \bibinfo {author} {\bibfnamefont {C.}~\bibnamefont {Monroe}},\ }\bibfield  {title} {\bibinfo {title} {Programmable quantum simulations of bosonic systems with trapped ions},\ }\href {https://doi.org/10.1103/PhysRevLett.131.033604} {\bibfield  {journal} {\bibinfo  {journal} {Phys. Rev. Lett.}\ }\textbf {\bibinfo {volume} {131}},\ \bibinfo {pages} {033604} (\bibinfo {year} {2023})}\BibitemShut {NoStop}%
\bibitem [{\citenamefont {Ding}\ \emph {et~al.}(2017)\citenamefont {Ding}, \citenamefont {Maslennikov}, \citenamefont {Habl\"utzel}, \citenamefont {Loh},\ and\ \citenamefont {Matsukevich}}]{parametric2017prl}%
  \BibitemOpen
  \bibfield  {author} {\bibinfo {author} {\bibfnamefont {S.}~\bibnamefont {Ding}}, \bibinfo {author} {\bibfnamefont {G.}~\bibnamefont {Maslennikov}}, \bibinfo {author} {\bibfnamefont {R.}~\bibnamefont {Habl\"utzel}}, \bibinfo {author} {\bibfnamefont {H.}~\bibnamefont {Loh}},\ and\ \bibinfo {author} {\bibfnamefont {D.}~\bibnamefont {Matsukevich}},\ }\bibfield  {title} {\bibinfo {title} {Quantum parametric oscillator with trapped ions},\ }\href {https://doi.org/10.1103/PhysRevLett.119.150404} {\bibfield  {journal} {\bibinfo  {journal} {Phys. Rev. Lett.}\ }\textbf {\bibinfo {volume} {119}},\ \bibinfo {pages} {150404} (\bibinfo {year} {2017})}\BibitemShut {NoStop}%
\bibitem [{\citenamefont {Burd}\ \emph {et~al.}(2021)\citenamefont {Burd}, \citenamefont {Srinivas}, \citenamefont {Knaack}, \citenamefont {Ge}, \citenamefont {Wilson}, \citenamefont {Wineland}, \citenamefont {Leibfried}, \citenamefont {Bollinger}, \citenamefont {Allcock},\ and\ \citenamefont {Slichter}}]{Burd2021}%
  \BibitemOpen
  \bibfield  {author} {\bibinfo {author} {\bibfnamefont {S.~C.}\ \bibnamefont {Burd}}, \bibinfo {author} {\bibfnamefont {R.}~\bibnamefont {Srinivas}}, \bibinfo {author} {\bibfnamefont {H.~M.}\ \bibnamefont {Knaack}}, \bibinfo {author} {\bibfnamefont {W.}~\bibnamefont {Ge}}, \bibinfo {author} {\bibfnamefont {A.~C.}\ \bibnamefont {Wilson}}, \bibinfo {author} {\bibfnamefont {D.~J.}\ \bibnamefont {Wineland}}, \bibinfo {author} {\bibfnamefont {D.}~\bibnamefont {Leibfried}}, \bibinfo {author} {\bibfnamefont {J.~J.}\ \bibnamefont {Bollinger}}, \bibinfo {author} {\bibfnamefont {D.~T.~C.}\ \bibnamefont {Allcock}},\ and\ \bibinfo {author} {\bibfnamefont {D.~H.}\ \bibnamefont {Slichter}},\ }\bibfield  {title} {\bibinfo {title} {Quantum amplification of boson-mediated interactions},\ }\href {https://doi.org/10.1038/s41567-021-01237-9} {\bibfield  {journal} {\bibinfo  {journal} {Nat. Phys.}\ }\textbf {\bibinfo {volume} {17}},\ \bibinfo {pages} {898} (\bibinfo {year} {2021})}\BibitemShut {NoStop}%
\bibitem [{\citenamefont {Hou}\ \emph {et~al.}(2024)\citenamefont {Hou}, \citenamefont {Wu}, \citenamefont {Erickson}, \citenamefont {Zarantonello}, \citenamefont {Brandt}, \citenamefont {Cole}, \citenamefont {Wilson}, \citenamefont {Slichter},\ and\ \citenamefont {Leibfried}}]{Leibfried2024prx}%
  \BibitemOpen
  \bibfield  {author} {\bibinfo {author} {\bibfnamefont {P.-Y.}\ \bibnamefont {Hou}}, \bibinfo {author} {\bibfnamefont {J.~J.}\ \bibnamefont {Wu}}, \bibinfo {author} {\bibfnamefont {S.~D.}\ \bibnamefont {Erickson}}, \bibinfo {author} {\bibfnamefont {G.}~\bibnamefont {Zarantonello}}, \bibinfo {author} {\bibfnamefont {A.~D.}\ \bibnamefont {Brandt}}, \bibinfo {author} {\bibfnamefont {D.~C.}\ \bibnamefont {Cole}}, \bibinfo {author} {\bibfnamefont {A.~C.}\ \bibnamefont {Wilson}}, \bibinfo {author} {\bibfnamefont {D.~H.}\ \bibnamefont {Slichter}},\ and\ \bibinfo {author} {\bibfnamefont {D.}~\bibnamefont {Leibfried}},\ }\bibfield  {title} {\bibinfo {title} {Indirect cooling of weakly coupled trapped-ion mechanical oscillators},\ }\href {https://doi.org/10.1103/PhysRevX.14.021003} {\bibfield  {journal} {\bibinfo  {journal} {Phys. Rev. X}\ }\textbf {\bibinfo {volume} {14}},\ \bibinfo {pages} {021003} (\bibinfo {year} {2024})}\BibitemShut {NoStop}%
\bibitem [{\citenamefont {B{\v{a}}z{\v{a}}van}\ \emph {et~al.}(2024)\citenamefont {B{\v{a}}z{\v{a}}van}, \citenamefont {Saner}, \citenamefont {Tirrito}, \citenamefont {Araneda}, \citenamefont {Srinivas},\ and\ \citenamefont {Bermudez}}]{Bazavan2024commphys}%
  \BibitemOpen
  \bibfield  {author} {\bibinfo {author} {\bibfnamefont {O.}~\bibnamefont {B{\v{a}}z{\v{a}}van}}, \bibinfo {author} {\bibfnamefont {S.}~\bibnamefont {Saner}}, \bibinfo {author} {\bibfnamefont {E.}~\bibnamefont {Tirrito}}, \bibinfo {author} {\bibfnamefont {G.}~\bibnamefont {Araneda}}, \bibinfo {author} {\bibfnamefont {R.}~\bibnamefont {Srinivas}},\ and\ \bibinfo {author} {\bibfnamefont {A.}~\bibnamefont {Bermudez}},\ }\bibfield  {title} {\bibinfo {title} {Synthetic $\mathbb{Z}_2$ gauge theories based on parametric excitations of trapped ions},\ }\href {https://doi.org/10.1038/s42005-024-01691-w} {\bibfield  {journal} {\bibinfo  {journal} {Communications Physics}\ }\textbf {\bibinfo {volume} {7}},\ \bibinfo {pages} {229} (\bibinfo {year} {2024})}\BibitemShut {NoStop}%
\bibitem [{\citenamefont {Bermudez}\ \emph {et~al.}(2013)\citenamefont {Bermudez}, \citenamefont {Bruderer},\ and\ \citenamefont {Plenio}}]{Bermudez2013prl}%
  \BibitemOpen
  \bibfield  {author} {\bibinfo {author} {\bibfnamefont {A.}~\bibnamefont {Bermudez}}, \bibinfo {author} {\bibfnamefont {M.}~\bibnamefont {Bruderer}},\ and\ \bibinfo {author} {\bibfnamefont {M.~B.}\ \bibnamefont {Plenio}},\ }\bibfield  {title} {\bibinfo {title} {Controlling and measuring quantum transport of heat in trapped-ion crystals},\ }\href {https://doi.org/10.1103/PhysRevLett.111.040601} {\bibfield  {journal} {\bibinfo  {journal} {Phys. Rev. Lett.}\ }\textbf {\bibinfo {volume} {111}},\ \bibinfo {pages} {040601} (\bibinfo {year} {2013})}\BibitemShut {NoStop}%
\bibitem [{\citenamefont {Porras}\ and\ \citenamefont {Fern{\'a}ndez-Lorenzo}(2019)}]{porras2019topological}%
  \BibitemOpen
  \bibfield  {author} {\bibinfo {author} {\bibfnamefont {D.}~\bibnamefont {Porras}}\ and\ \bibinfo {author} {\bibfnamefont {S.}~\bibnamefont {Fern{\'a}ndez-Lorenzo}},\ }\bibfield  {title} {\bibinfo {title} {Topological amplification in photonic lattices},\ }\href {https://journals.aps.org/prl/abstract/10.1103/PhysRevLett.122.143901} {\bibfield  {journal} {\bibinfo  {journal} {Phys. Rev. Lett.}\ }\textbf {\bibinfo {volume} {122}},\ \bibinfo {pages} {143901} (\bibinfo {year} {2019})}\BibitemShut {NoStop}%
\bibitem [{\citenamefont {Wanjura}\ \emph {et~al.}(2020)\citenamefont {Wanjura}, \citenamefont {Brunelli},\ and\ \citenamefont {Nunnenkamp}}]{wanjura2020topological}%
  \BibitemOpen
  \bibfield  {author} {\bibinfo {author} {\bibfnamefont {C.~C.}\ \bibnamefont {Wanjura}}, \bibinfo {author} {\bibfnamefont {M.}~\bibnamefont {Brunelli}},\ and\ \bibinfo {author} {\bibfnamefont {A.}~\bibnamefont {Nunnenkamp}},\ }\bibfield  {title} {\bibinfo {title} {Topological framework for directional amplification in driven-dissipative cavity arrays},\ }\href {https://www.nature.com/articles/s41467-020-16863-9} {\bibfield  {journal} {\bibinfo  {journal} {Nature communications}\ }\textbf {\bibinfo {volume} {11}},\ \bibinfo {pages} {3149} (\bibinfo {year} {2020})}\BibitemShut {NoStop}%
\bibitem [{\citenamefont {Ramos}\ \emph {et~al.}(2021)\citenamefont {Ramos}, \citenamefont {Garc{\'\i}a-Ripoll},\ and\ \citenamefont {Porras}}]{ramos2021topological}%
  \BibitemOpen
  \bibfield  {author} {\bibinfo {author} {\bibfnamefont {T.}~\bibnamefont {Ramos}}, \bibinfo {author} {\bibfnamefont {J.~J.}\ \bibnamefont {Garc{\'\i}a-Ripoll}},\ and\ \bibinfo {author} {\bibfnamefont {D.}~\bibnamefont {Porras}},\ }\bibfield  {title} {\bibinfo {title} {Topological input-output theory for directional amplification},\ }\href {https://journals.aps.org/pra/abstract/10.1103/PhysRevA.103.033513} {\bibfield  {journal} {\bibinfo  {journal} {Phys. Rev. A}\ }\textbf {\bibinfo {volume} {103}},\ \bibinfo {pages} {033513} (\bibinfo {year} {2021})}\BibitemShut {NoStop}%
\bibitem [{\citenamefont {G{\'o}mez-Le{\'o}n}\ \emph {et~al.}(2023)\citenamefont {G{\'o}mez-Le{\'o}n}, \citenamefont {Ramos}, \citenamefont {Gonz{\'a}lez-Tudela},\ and\ \citenamefont {Porras}}]{gomez2023driven}%
  \BibitemOpen
  \bibfield  {author} {\bibinfo {author} {\bibfnamefont {{\'A}.}~\bibnamefont {G{\'o}mez-Le{\'o}n}}, \bibinfo {author} {\bibfnamefont {T.}~\bibnamefont {Ramos}}, \bibinfo {author} {\bibfnamefont {A.}~\bibnamefont {Gonz{\'a}lez-Tudela}},\ and\ \bibinfo {author} {\bibfnamefont {D.}~\bibnamefont {Porras}},\ }\bibfield  {title} {\bibinfo {title} {Driven-dissipative topological phases in parametric resonator arrays},\ }\href {https://arxiv.org/abs/2207.13715} {\bibfield  {journal} {\bibinfo  {journal} {Quantum}\ }\textbf {\bibinfo {volume} {7}},\ \bibinfo {pages} {1016} (\bibinfo {year} {2023})}\BibitemShut {NoStop}%
\bibitem [{\citenamefont {McDonald}\ \emph {et~al.}(2018)\citenamefont {McDonald}, \citenamefont {Pereg-Barnea},\ and\ \citenamefont {Clerk}}]{mcdonald2018phase}%
  \BibitemOpen
  \bibfield  {author} {\bibinfo {author} {\bibfnamefont {A.}~\bibnamefont {McDonald}}, \bibinfo {author} {\bibfnamefont {T.}~\bibnamefont {Pereg-Barnea}},\ and\ \bibinfo {author} {\bibfnamefont {A.}~\bibnamefont {Clerk}},\ }\bibfield  {title} {\bibinfo {title} {Phase-dependent chiral transport and effective non-hermitian dynamics in a bosonic kitaev-majorana chain},\ }\href {https://journals.aps.org/prx/abstract/10.1103/PhysRevX.8.041031} {\bibfield  {journal} {\bibinfo  {journal} {Phys. Rev. X}\ }\textbf {\bibinfo {volume} {8}},\ \bibinfo {pages} {041031} (\bibinfo {year} {2018})}\BibitemShut {NoStop}%
\bibitem [{\citenamefont {Busnaina}\ \emph {et~al.}(2024)\citenamefont {Busnaina}, \citenamefont {Shi}, \citenamefont {McDonald}, \citenamefont {Dubyna}, \citenamefont {Nsanzineza}, \citenamefont {Hung}, \citenamefont {Chang}, \citenamefont {Clerk},\ and\ \citenamefont {Wilson}}]{busnaina2024quantum}%
  \BibitemOpen
  \bibfield  {author} {\bibinfo {author} {\bibfnamefont {J.~H.}\ \bibnamefont {Busnaina}}, \bibinfo {author} {\bibfnamefont {Z.}~\bibnamefont {Shi}}, \bibinfo {author} {\bibfnamefont {A.}~\bibnamefont {McDonald}}, \bibinfo {author} {\bibfnamefont {D.}~\bibnamefont {Dubyna}}, \bibinfo {author} {\bibfnamefont {I.}~\bibnamefont {Nsanzineza}}, \bibinfo {author} {\bibfnamefont {J.~S.}\ \bibnamefont {Hung}}, \bibinfo {author} {\bibfnamefont {C.~S.}\ \bibnamefont {Chang}}, \bibinfo {author} {\bibfnamefont {A.~A.}\ \bibnamefont {Clerk}},\ and\ \bibinfo {author} {\bibfnamefont {C.~M.}\ \bibnamefont {Wilson}},\ }\bibfield  {title} {\bibinfo {title} {Quantum simulation of the bosonic kitaev chain},\ }\href {https://www.nature.com/articles/s41467-024-47186-8} {\bibfield  {journal} {\bibinfo  {journal} {Nature Communications}\ }\textbf {\bibinfo {volume} {15}},\ \bibinfo {pages} {3065} (\bibinfo {year} {2024})}\BibitemShut {NoStop}%
\bibitem [{\citenamefont {Slim}\ \emph {et~al.}(2024)\citenamefont {Slim}, \citenamefont {Wanjura}, \citenamefont {Brunelli}, \citenamefont {Del~Pino}, \citenamefont {Nunnenkamp},\ and\ \citenamefont {Verhagen}}]{slim2024optomechanical}%
  \BibitemOpen
  \bibfield  {author} {\bibinfo {author} {\bibfnamefont {J.~J.}\ \bibnamefont {Slim}}, \bibinfo {author} {\bibfnamefont {C.~C.}\ \bibnamefont {Wanjura}}, \bibinfo {author} {\bibfnamefont {M.}~\bibnamefont {Brunelli}}, \bibinfo {author} {\bibfnamefont {J.}~\bibnamefont {Del~Pino}}, \bibinfo {author} {\bibfnamefont {A.}~\bibnamefont {Nunnenkamp}},\ and\ \bibinfo {author} {\bibfnamefont {E.}~\bibnamefont {Verhagen}},\ }\bibfield  {title} {\bibinfo {title} {Optomechanical realization of the bosonic kitaev chain},\ }\href {https://www.nature.com/articles/s41586-024-07174-w} {\bibfield  {journal} {\bibinfo  {journal} {Nature}\ }\textbf {\bibinfo {volume} {627}},\ \bibinfo {pages} {767} (\bibinfo {year} {2024})}\BibitemShut {NoStop}%
\bibitem [{\citenamefont {Pagano}\ \emph {et~al.}(2018)\citenamefont {Pagano}, \citenamefont {Hess}, \citenamefont {Kaplan}, \citenamefont {Tan}, \citenamefont {Richerme}, \citenamefont {Becker}, \citenamefont {Kyprianidis}, \citenamefont {Zhang}, \citenamefont {Birckelbaw}, \citenamefont {Hernandez}, \citenamefont {Wu},\ and\ \citenamefont {Monroe}}]{Pagano2019}%
  \BibitemOpen
  \bibfield  {author} {\bibinfo {author} {\bibfnamefont {G.}~\bibnamefont {Pagano}}, \bibinfo {author} {\bibfnamefont {P.~W.}\ \bibnamefont {Hess}}, \bibinfo {author} {\bibfnamefont {H.~B.}\ \bibnamefont {Kaplan}}, \bibinfo {author} {\bibfnamefont {W.~L.}\ \bibnamefont {Tan}}, \bibinfo {author} {\bibfnamefont {P.}~\bibnamefont {Richerme}}, \bibinfo {author} {\bibfnamefont {P.}~\bibnamefont {Becker}}, \bibinfo {author} {\bibfnamefont {A.}~\bibnamefont {Kyprianidis}}, \bibinfo {author} {\bibfnamefont {J.}~\bibnamefont {Zhang}}, \bibinfo {author} {\bibfnamefont {E.}~\bibnamefont {Birckelbaw}}, \bibinfo {author} {\bibfnamefont {M.~R.}\ \bibnamefont {Hernandez}}, \bibinfo {author} {\bibfnamefont {Y.}~\bibnamefont {Wu}},\ and\ \bibinfo {author} {\bibfnamefont {C.}~\bibnamefont {Monroe}},\ }\bibfield  {title} {\bibinfo {title} {Cryogenic trapped-ion system for large scale quantum simulation},\ }\href {https://doi.org/10.1088/2058-9565/aae0fe} {\bibfield  {journal} {\bibinfo  {journal} {Quantum Science and
  Technology}\ }\textbf {\bibinfo {volume} {4}},\ \bibinfo {pages} {014004} (\bibinfo {year} {2018})}\BibitemShut {NoStop}%
\bibitem [{\citenamefont {James}(1997)}]{james1997quantum}%
  \BibitemOpen
  \bibfield  {author} {\bibinfo {author} {\bibfnamefont {D.~F.}\ \bibnamefont {James}},\ }\href {https://books.google.es/books?hl=es&lr=&id=t3BqDQAAQBAJ&oi=fnd&pg=PA345&dq=quantum+dynamics+of+cold+tr&ots=HQiX2Vbg6Y&sig=g8Pq8LRV8yjA2b0_RRcz-dP1_e8#v=onepage&q&f=false} {\emph {\bibinfo {title} {Quantum dynamics of cold trapped ions with application to quantum computation}}},\ \bibinfo {type} {Tech. Rep.}\ (\bibinfo {year} {1997})\BibitemShut {NoStop}%
\bibitem [{\citenamefont {G{\'o}mez-Le{\'o}n}\ \emph {et~al.}(2022)\citenamefont {G{\'o}mez-Le{\'o}n}, \citenamefont {Ramos}, \citenamefont {Gonz{\'a}lez-Tudela},\ and\ \citenamefont {Porras}}]{gomez2022bridging}%
  \BibitemOpen
  \bibfield  {author} {\bibinfo {author} {\bibfnamefont {{\'A}.}~\bibnamefont {G{\'o}mez-Le{\'o}n}}, \bibinfo {author} {\bibfnamefont {T.}~\bibnamefont {Ramos}}, \bibinfo {author} {\bibfnamefont {A.}~\bibnamefont {Gonz{\'a}lez-Tudela}},\ and\ \bibinfo {author} {\bibfnamefont {D.}~\bibnamefont {Porras}},\ }\bibfield  {title} {\bibinfo {title} {Bridging the gap between topological non-hermitian physics and open quantum systems},\ }\href {https://journals.aps.org/pra/abstract/10.1103/PhysRevA.106.L011501} {\bibfield  {journal} {\bibinfo  {journal} {Phys. Rev. A}\ }\textbf {\bibinfo {volume} {106}},\ \bibinfo {pages} {L011501} (\bibinfo {year} {2022})}\BibitemShut {NoStop}%
\bibitem [{\citenamefont {Gardiner}\ and\ \citenamefont {Zoller}(2004)}]{gardiner_book}%
  \BibitemOpen
  \bibfield  {author} {\bibinfo {author} {\bibfnamefont {C.}~\bibnamefont {Gardiner}}\ and\ \bibinfo {author} {\bibfnamefont {P.}~\bibnamefont {Zoller}},\ }\href {https://books.google.es/books?hl=es&lr=&id=a_xsT8oGhdgC&oi=fnd&pg=PA1&dq=Quantum+noise:+a+handbook&ots=k0s_pPcWwa&sig=HZvrE19w-bQCrcn8eZmboUATf3I#v=onepage&q=Quantum%20noise%3A%20a%20handbook&f=false} {\emph {\bibinfo {title} {Quantum noise: a handbook of Markovian and non-Markovian quantum stochastic methods with applications to quantum optics}}}\ (\bibinfo  {publisher} {Springer Science \& Business Media},\ \bibinfo {year} {2004})\BibitemShut {NoStop}%
\bibitem [{\citenamefont {Ughrelidze}\ \emph {et~al.}(2024)\citenamefont {Ughrelidze}, \citenamefont {Flynn}, \citenamefont {Cobanera},\ and\ \citenamefont {Viola}}]{ughrelidze2024interplay}%
  \BibitemOpen
  \bibfield  {author} {\bibinfo {author} {\bibfnamefont {M.}~\bibnamefont {Ughrelidze}}, \bibinfo {author} {\bibfnamefont {V.}~\bibnamefont {Flynn}}, \bibinfo {author} {\bibfnamefont {E.}~\bibnamefont {Cobanera}},\ and\ \bibinfo {author} {\bibfnamefont {L.}~\bibnamefont {Viola}},\ }\bibfield  {title} {\bibinfo {title} {The interplay of finite and infinite size stability in quadratic bosonic lindbladians},\ }\href {https://journals.aps.org/pra/abstract/10.1103/PhysRevA.110.032207} {\bibfield  {journal} {\bibinfo  {journal} {Bull. Am. Phys. Soc.}\ } (\bibinfo {year} {2024})}\BibitemShut {NoStop}%
\bibitem [{\citenamefont {Okuma}\ and\ \citenamefont {Sato}(2023)}]{Sato2023review}%
  \BibitemOpen
  \bibfield  {author} {\bibinfo {author} {\bibfnamefont {N.}~\bibnamefont {Okuma}}\ and\ \bibinfo {author} {\bibfnamefont {M.}~\bibnamefont {Sato}},\ }\bibfield  {title} {\bibinfo {title} {Non-hermitian topological phenomena: A review},\ }\href {https://doi.org/https://doi.org/10.1146/annurev-conmatphys-040521-033133} {\bibfield  {journal} {\bibinfo  {journal} {Annual Review of Condensed Matter Physics}\ }\textbf {\bibinfo {volume} {14}},\ \bibinfo {pages} {83} (\bibinfo {year} {2023})}\BibitemShut {NoStop}%
\bibitem [{\citenamefont {Hasan}\ and\ \citenamefont {Kane}(2010)}]{hasan2010colloquium}%
  \BibitemOpen
  \bibfield  {author} {\bibinfo {author} {\bibfnamefont {M.~Z.}\ \bibnamefont {Hasan}}\ and\ \bibinfo {author} {\bibfnamefont {C.~L.}\ \bibnamefont {Kane}},\ }\bibfield  {title} {\bibinfo {title} {Colloquium: topological insulators},\ }\href {https://journals.aps.org/rmp/abstract/10.1103/RevModPhys.82.3045} {\bibfield  {journal} {\bibinfo  {journal} {Rev. Mod. Phys.}\ }\textbf {\bibinfo {volume} {82}},\ \bibinfo {pages} {3045} (\bibinfo {year} {2010})}\BibitemShut {NoStop}%
\bibitem [{\citenamefont {Ivanov}\ \emph {et~al.}(2016)\citenamefont {Ivanov}, \citenamefont {Vitanov},\ and\ \citenamefont {Singer}}]{ivanov2016high}%
  \BibitemOpen
  \bibfield  {author} {\bibinfo {author} {\bibfnamefont {P.~A.}\ \bibnamefont {Ivanov}}, \bibinfo {author} {\bibfnamefont {N.~V.}\ \bibnamefont {Vitanov}},\ and\ \bibinfo {author} {\bibfnamefont {K.}~\bibnamefont {Singer}},\ }\bibfield  {title} {\bibinfo {title} {High-precision force sensing using a single trapped ion},\ }\href {https://www.nature.com/articles/srep28078} {\bibfield  {journal} {\bibinfo  {journal} {Sci. Rep.}\ }\textbf {\bibinfo {volume} {6}},\ \bibinfo {pages} {28078} (\bibinfo {year} {2016})}\BibitemShut {NoStop}%
\bibitem [{\citenamefont {Di~Candia}\ \emph {et~al.}(2023)\citenamefont {Di~Candia}, \citenamefont {Minganti}, \citenamefont {Petrovnin}, \citenamefont {Paraoanu},\ and\ \citenamefont {Felicetti}}]{di2023critical}%
  \BibitemOpen
  \bibfield  {author} {\bibinfo {author} {\bibfnamefont {R.}~\bibnamefont {Di~Candia}}, \bibinfo {author} {\bibfnamefont {F.}~\bibnamefont {Minganti}}, \bibinfo {author} {\bibfnamefont {K.}~\bibnamefont {Petrovnin}}, \bibinfo {author} {\bibfnamefont {G.}~\bibnamefont {Paraoanu}},\ and\ \bibinfo {author} {\bibfnamefont {S.}~\bibnamefont {Felicetti}},\ }\bibfield  {title} {\bibinfo {title} {Critical parametric quantum sensing},\ }\href {https://www.nature.com/articles/s41534-023-00690-z} {\bibfield  {journal} {\bibinfo  {journal} {npj Quantum Inf.}\ }\textbf {\bibinfo {volume} {9}},\ \bibinfo {pages} {23} (\bibinfo {year} {2023})}\BibitemShut {NoStop}%
\bibitem [{\citenamefont {Streed}\ \emph {et~al.}(2011)\citenamefont {Streed}, \citenamefont {Norton}, \citenamefont {Jechow}, \citenamefont {Weinhold},\ and\ \citenamefont {Kielpinski}}]{streed2011imaging}%
  \BibitemOpen
  \bibfield  {author} {\bibinfo {author} {\bibfnamefont {E.~W.}\ \bibnamefont {Streed}}, \bibinfo {author} {\bibfnamefont {B.~G.}\ \bibnamefont {Norton}}, \bibinfo {author} {\bibfnamefont {A.}~\bibnamefont {Jechow}}, \bibinfo {author} {\bibfnamefont {T.~J.}\ \bibnamefont {Weinhold}},\ and\ \bibinfo {author} {\bibfnamefont {D.}~\bibnamefont {Kielpinski}},\ }\bibfield  {title} {\bibinfo {title} {Imaging of trapped ions with a microfabricated optic for quantum information processing},\ }\href {https://journals.aps.org/prl/abstract/10.1103/PhysRevLett.106.010502} {\bibfield  {journal} {\bibinfo  {journal} {Phys. Rev. Lett.}\ }\textbf {\bibinfo {volume} {106}},\ \bibinfo {pages} {010502} (\bibinfo {year} {2011})}\BibitemShut {NoStop}%
\bibitem [{\citenamefont {Drechsler}\ \emph {et~al.}(2021)\citenamefont {Drechsler}, \citenamefont {Wolf}, \citenamefont {Schmiegelow},\ and\ \citenamefont {Schmidt-Kaler}}]{drechsler2021optical}%
  \BibitemOpen
  \bibfield  {author} {\bibinfo {author} {\bibfnamefont {M.}~\bibnamefont {Drechsler}}, \bibinfo {author} {\bibfnamefont {S.}~\bibnamefont {Wolf}}, \bibinfo {author} {\bibfnamefont {C.~T.}\ \bibnamefont {Schmiegelow}},\ and\ \bibinfo {author} {\bibfnamefont {F.}~\bibnamefont {Schmidt-Kaler}},\ }\bibfield  {title} {\bibinfo {title} {Optical superresolution sensing of a trapped ion’s wave packet size},\ }\href {https://journals.aps.org/prl/abstract/10.1103/PhysRevLett.127.143602} {\bibfield  {journal} {\bibinfo  {journal} {Phys. Rev. Lett.}\ }\textbf {\bibinfo {volume} {127}},\ \bibinfo {pages} {143602} (\bibinfo {year} {2021})}\BibitemShut {NoStop}%
\bibitem [{\citenamefont {Hasse}\ \emph {et~al.}(2024)\citenamefont {Hasse}, \citenamefont {Palani}, \citenamefont {Thomm}, \citenamefont {Warring},\ and\ \citenamefont {Schaetz}}]{hasse2024phase}%
  \BibitemOpen
  \bibfield  {author} {\bibinfo {author} {\bibfnamefont {F.}~\bibnamefont {Hasse}}, \bibinfo {author} {\bibfnamefont {D.}~\bibnamefont {Palani}}, \bibinfo {author} {\bibfnamefont {R.}~\bibnamefont {Thomm}}, \bibinfo {author} {\bibfnamefont {U.}~\bibnamefont {Warring}},\ and\ \bibinfo {author} {\bibfnamefont {T.}~\bibnamefont {Schaetz}},\ }\bibfield  {title} {\bibinfo {title} {Phase-stable traveling waves stroboscopically matched for superresolved observation of trapped-ion dynamics},\ }\href {https://journals.aps.org/pra/abstract/10.1103/PhysRevA.109.053105} {\bibfield  {journal} {\bibinfo  {journal} {Phys. Rev. A}\ }\textbf {\bibinfo {volume} {109}},\ \bibinfo {pages} {053105} (\bibinfo {year} {2024})}\BibitemShut {NoStop}%
\bibitem [{\citenamefont {Degen}\ \emph {et~al.}(2017)\citenamefont {Degen}, \citenamefont {Reinhard},\ and\ \citenamefont {Cappellaro}}]{degen2017quantum}%
  \BibitemOpen
  \bibfield  {author} {\bibinfo {author} {\bibfnamefont {C.~L.}\ \bibnamefont {Degen}}, \bibinfo {author} {\bibfnamefont {F.}~\bibnamefont {Reinhard}},\ and\ \bibinfo {author} {\bibfnamefont {P.}~\bibnamefont {Cappellaro}},\ }\bibfield  {title} {\bibinfo {title} {Quantum sensing},\ }\href {https://journals.aps.org/rmp/abstract/10.1103/RevModPhys.89.035002} {\bibfield  {journal} {\bibinfo  {journal} {Rev. Mod. Phys.}\ }\textbf {\bibinfo {volume} {89}},\ \bibinfo {pages} {035002} (\bibinfo {year} {2017})}\BibitemShut {NoStop}%
\bibitem [{\citenamefont {Liang}\ \emph {et~al.}(2023)\citenamefont {Liang}, \citenamefont {Zhu}, \citenamefont {He}, \citenamefont {Chen}, \citenamefont {Wang}, \citenamefont {Li}, \citenamefont {Fu}, \citenamefont {Gao}, \citenamefont {Chen}, \citenamefont {Li} \emph {et~al.}}]{liang2023yoctonewton}%
  \BibitemOpen
  \bibfield  {author} {\bibinfo {author} {\bibfnamefont {T.}~\bibnamefont {Liang}}, \bibinfo {author} {\bibfnamefont {S.}~\bibnamefont {Zhu}}, \bibinfo {author} {\bibfnamefont {P.}~\bibnamefont {He}}, \bibinfo {author} {\bibfnamefont {Z.}~\bibnamefont {Chen}}, \bibinfo {author} {\bibfnamefont {Y.}~\bibnamefont {Wang}}, \bibinfo {author} {\bibfnamefont {C.}~\bibnamefont {Li}}, \bibinfo {author} {\bibfnamefont {Z.}~\bibnamefont {Fu}}, \bibinfo {author} {\bibfnamefont {X.}~\bibnamefont {Gao}}, \bibinfo {author} {\bibfnamefont {X.}~\bibnamefont {Chen}}, \bibinfo {author} {\bibfnamefont {N.}~\bibnamefont {Li}}, \emph {et~al.},\ }\bibfield  {title} {\bibinfo {title} {Yoctonewton force detection based on optically levitated oscillator},\ }\href {https://www.sciencedirect.com/science/article/pii/S2667325822003879} {\bibfield  {journal} {\bibinfo  {journal} {Fundam. Res.}\ }\textbf {\bibinfo {volume} {3}},\ \bibinfo {pages} {57} (\bibinfo {year} {2023})}\BibitemShut {NoStop}%
\bibitem [{\citenamefont {Maiwald}\ \emph {et~al.}(2009)\citenamefont {Maiwald}, \citenamefont {Leibfried}, \citenamefont {Britton}, \citenamefont {Bergquist}, \citenamefont {Leuchs},\ and\ \citenamefont {Wineland}}]{maiwald2009stylus}%
  \BibitemOpen
  \bibfield  {author} {\bibinfo {author} {\bibfnamefont {R.}~\bibnamefont {Maiwald}}, \bibinfo {author} {\bibfnamefont {D.}~\bibnamefont {Leibfried}}, \bibinfo {author} {\bibfnamefont {J.}~\bibnamefont {Britton}}, \bibinfo {author} {\bibfnamefont {J.~C.}\ \bibnamefont {Bergquist}}, \bibinfo {author} {\bibfnamefont {G.}~\bibnamefont {Leuchs}},\ and\ \bibinfo {author} {\bibfnamefont {D.~J.}\ \bibnamefont {Wineland}},\ }\bibfield  {title} {\bibinfo {title} {Stylus ion trap for enhanced access and sensing},\ }\href {https://www.nature.com/articles/nphys1311} {\bibfield  {journal} {\bibinfo  {journal} {Nat. Phys.}\ }\textbf {\bibinfo {volume} {5}},\ \bibinfo {pages} {551} (\bibinfo {year} {2009})}\BibitemShut {NoStop}%
\bibitem [{\citenamefont {Peano}\ \emph {et~al.}(2016)\citenamefont {Peano}, \citenamefont {Houde}, \citenamefont {Marquardt},\ and\ \citenamefont {Clerk}}]{peano2016topological}%
  \BibitemOpen
  \bibfield  {author} {\bibinfo {author} {\bibfnamefont {V.}~\bibnamefont {Peano}}, \bibinfo {author} {\bibfnamefont {M.}~\bibnamefont {Houde}}, \bibinfo {author} {\bibfnamefont {F.}~\bibnamefont {Marquardt}},\ and\ \bibinfo {author} {\bibfnamefont {A.~A.}\ \bibnamefont {Clerk}},\ }\bibfield  {title} {\bibinfo {title} {Topological quantum fluctuations and traveling wave amplifiers},\ }\href {https://journals.aps.org/prx/abstract/10.1103/PhysRevX.6.041026} {\bibfield  {journal} {\bibinfo  {journal} {Phys. Rev. X}\ }\textbf {\bibinfo {volume} {6}},\ \bibinfo {pages} {041026} (\bibinfo {year} {2016})}\BibitemShut {NoStop}%
\bibitem [{\citenamefont {Vega}\ \emph {et~al.}(2024)\citenamefont {Vega}, \citenamefont {de~las Heras}, \citenamefont {Porras},\ and\ \citenamefont {González-Tudela}}]{vega2024arXiv}%
  \BibitemOpen
  \bibfield  {author} {\bibinfo {author} {\bibfnamefont {C.}~\bibnamefont {Vega}}, \bibinfo {author} {\bibfnamefont {A.~M.}\ \bibnamefont {de~las Heras}}, \bibinfo {author} {\bibfnamefont {D.}~\bibnamefont {Porras}},\ and\ \bibinfo {author} {\bibfnamefont {A.}~\bibnamefont {González-Tudela}},\ }\href {https://arxiv.org/abs/2405.10176} {\bibinfo {title} {Topological, multi-mode amplification induced by non-reciprocal, long-range dissipative couplings}} (\bibinfo {year} {2024}),\ \Eprint {https://arxiv.org/abs/2405.10176} {arXiv:2405.10176 [quant-ph]} \BibitemShut {NoStop}%
\bibitem [{\citenamefont {Altland}\ and\ \citenamefont {Zirnbauer}(1997)}]{altland1997nonstandard}%
  \BibitemOpen
  \bibfield  {author} {\bibinfo {author} {\bibfnamefont {A.}~\bibnamefont {Altland}}\ and\ \bibinfo {author} {\bibfnamefont {M.~R.}\ \bibnamefont {Zirnbauer}},\ }\bibfield  {title} {\bibinfo {title} {Nonstandard symmetry classes in mesoscopic normal-superconducting hybrid structures},\ }\href {https://journals.aps.org/prb/abstract/10.1103/PhysRevB.55.1142} {\bibfield  {journal} {\bibinfo  {journal} {Phys. Rev. B}\ }\textbf {\bibinfo {volume} {55}},\ \bibinfo {pages} {1142} (\bibinfo {year} {1997})}\BibitemShut {NoStop}%
\end{thebibliography}%

\end{document}